# Food Productivity Trends from Hybrid Corn: Statistical Analysis of Patents and Field-test data

Mariam Barry[1]

*ENSTA ParisTech*

Giorgio Triulzi and Christopher L. Magee

*MIT Institute for Data, Systems, and Society*

## Abstract

In this research we study productivity trends of hybrid corn - an important subdomain of food production. We estimate the yearly rate of yield improvement of hybrid corn (measured as bushel per acre) by using both information on yields contained in US patent documents for patented hybrid corn varieties and on field-test data of several hybrid corn varieties performed at US State level. We have used a generalization of Moore's law to fit productivity trends and obtain the performance improvement rate by analyzing time series of hybrid corn performance for a period covering the last thirty years. The linear regressions results obtained from different data sources indicate that the estimated improvement rates are between 0.012 and 0.024 (1.2 to 2.4% improvement per year). In particular, using yields reported in a sample of patents filed between 1985 and 2010, we estimated an improvement rate of 0.015 ($R^2 = 0.74$, $P_{value} = 1.37 * 10^{-8}$). Moreover, we apply two predicting models developed by *Benson and Magee (2015)* and *Triulzi and Magee (2016)* that only use patent metadata to estimate the rate of improvement. We compare these predicted values to the rate estimated using US States' field-test data. We find that, due to a turning point in patenting practices which begun in 2008, only the predicted rate (rate = 0.015) using patents filed before 2008 is consistent with the empirical rate. Finally, we also investigate at the micro level - on the basis of 70 patents (granted between 1986 and 2015) - whether the number of citations received by a patent is correlated with performance achieved by the patented variety. We found that the relative performance (yield ratio) of the patented seed is positively correlated with the total number of citations received by the patent (until December 2015) but not the citations received within 3 years after the granted year, with the patent application year used as control variable.

**Keywords**

[1] Corresponding author: mariam.sa.barry@gmail.com

# 1. Introduction

Thomas Robert Malthus was one of the first scholars to study demographic growth and its interaction with food supply. He argued in his 1798 essay [1] that population will grow geometrically (exponentially) and overtake the growth in food production (which he quantifies as arithmetic, linear) leading to intermittent famines and poverty in the world. These conclusions have been followed by others, for example Lester R. Brown claimed in his 1991 paper [5] that growth in average world grain was dramatically slowing down. However, Malthus' theory has been largely criticized by many anti-Malthusians such as the American economist Julian L. Simon who said[2] that [2]: *"The ultimate resource is people-skilled, spirited, and hopeful people who will exert their wills and imaginations for their own benefits, and so, inevitably, for the benefits of us all."* [3]. More than two hundred years later, we can say that Malthus' catastrophic predictions were not correct, or at least for a worldwide point of view[3]. Indeed, agricultural changes have increased world capacity to provide food in quantity, especially for developed countries. The FAO (Food and Agricultural Organization of the United Nations) showed in a 2012 report that food supply has grown faster than population [4].

Over time, innovations in irrigation techniques coupled with an expansion of arable lands and important rise in crops yields have resulted in large increases in food productivity in the world. The Green Revolution, initially appearing in the US at the end of 1960s [2] and which refers to the increase in crops production by using modern techniques and agricultural technologies such as fertilizers and pesticides, combined with mechanization in farming, have enhanced the global food supply. *Evenson and Gollin (2004) [6]* showed that without the Green Revolution, crop yield in developing countries would have been 20% to 24% lower. These modern agricultural techniques and changes have been followed by the introduction of Biotechnology and development of GMO (Genetically Modified Organisms – which seeds are created in labs as result of high-tech and sophisticated techniques like gene-splicing [7] )and hybrid crops. Since its introduction, Biotech crops have been widely used and substituted to former techniques in many countries. According to an article [8], for a variety of specific cases, biotech crops reach 90% substitution only 14 years after introduction (as shown on Figure 2).

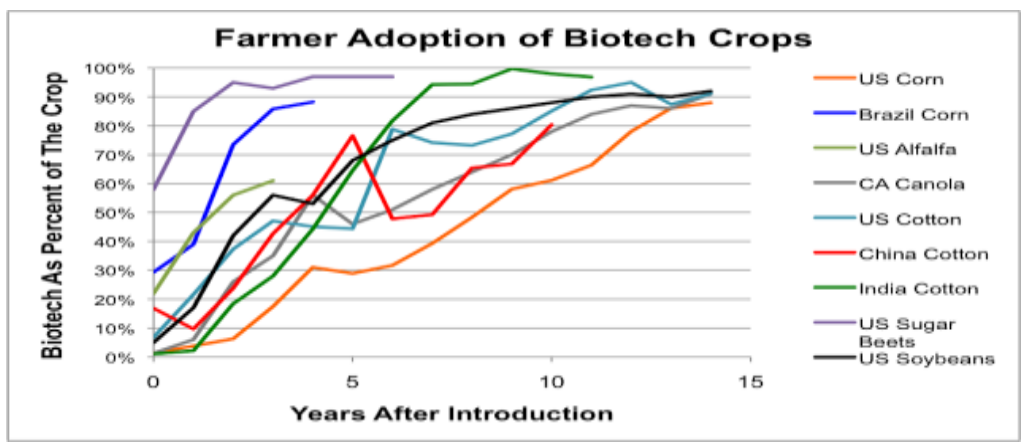

*Figure 1: Biotech crops adoption by farmers*

---

[2] See Julian Lincoln Simon 's book "The ultimate Resource " (1981, p. 348).
[3] Note that many developing countries are still facing hunger issues, for example in 2014, 176 Millions people were undernourished in Southern Asia countries and 23.2% of prevalence in Sub-Saharan African. Source : FAO et al, 2015 p. 8-12.

In particular, US corn, soybeans, sugar beets and cotton exceeded 85% of bio crops 14 years after bio crops introduction while China and India Cotton and Brazil Corn bio crops represented more than 80% of crops only 10 years after Bio crops introduction.

The United States is the first producer and consumer of corn in the world, with 13 601 Million bushels of corns produced in the last marketing year (October 1st to September 30, 2015) and 11 870 million bushel consumed [9-10] according to USDA (United States Department of agriculture) data. Measured in bushels/acre, the United States corn productivity average has more than quintupled from 1940 to 2015. These performances can be explained, as mentioned below, by improvements and changes in agricultural techniques and by the diffusion of the 'Green Revolution', i.e. of the systematic use of pesticides, herbicides and fertilizers. However, the introduction of hybrid corn in the 1930s also played a crucial role. Indeed, only 30 years after hybrid corn seed had become available for American farmers, around **95%** of United States acreage were planted in hybrid corn and compared to 1930, more than **20%** more corn was produced on **25%** fewer acres [11]. Figure 2 below shows the main years in which we observed important progress achieved in U.S. corn [23].

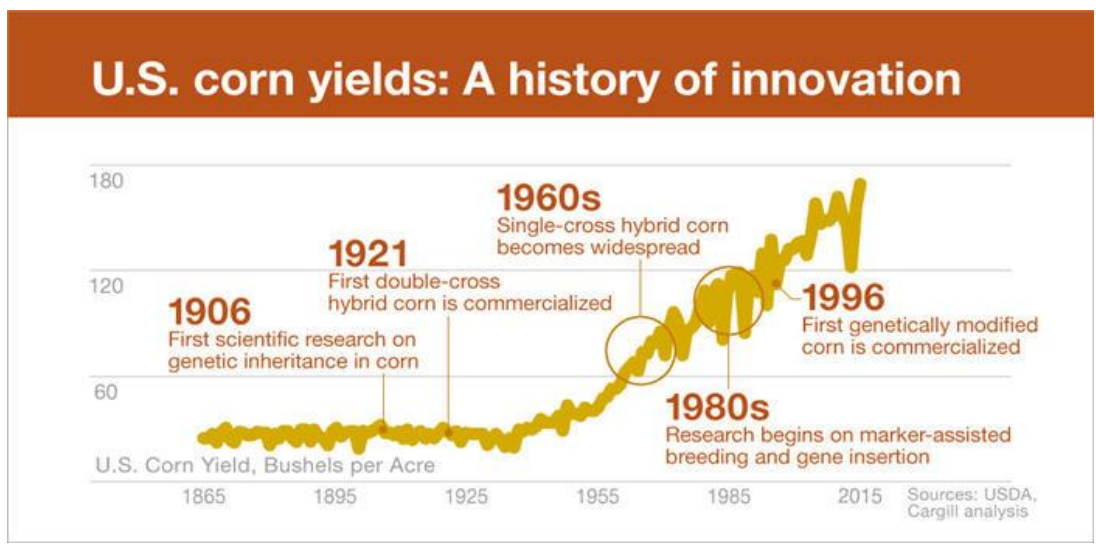

*Figure 2: History and Key Years in US Corn innovation_Source : USDA data, by Cargill analysis*

Note that corn (thus hybrid corn) represents an important branch in food production since about 80% of corn produced is use in farming for animal feeding. For example, 1 hectare of corn produces about 9000 pounds of meat. Corn is largely used by food processing industries for its starch concentration.

A goal of this research is to quantify the rate of progress in hybrid corn productivity which is an important subdomain in food production. In particular, hybrid corn is an interesting case study because first, it represents one of the inventions that subverted Malthus catastrophic predictions and secondly it allows a researcher to link macro data on corn yields with micro data at the seed level by using patents and field-test data. This leads us to the first research question investigated in this report, i.e. *what is the observed rate of improvement in bushels of hybrid corn per acre in the United States*? Moreover, the hybrid corn domain is one of the rare cases where patents refers to one product (hybrid seed) and include in particular performance data on the patented product. This makes hybrid corn a unique case study of the speed of technological change. Technological change is known to be cumulative, i.e. build on previously produced knowledge.

Citations in patent documents are an important indicator to quantify this effect. In that context and for a micro level, the second research question analyzed in this paper is: *do better performing hybrid corn varieties receive more citations (i.e. follow-on inventions)*?

This paper proceeds as follows. Section 2 gives an overview on previous and more recent literature on performance trends and technological progress followed by section 3, where we detail the data used to conduct this research and how we collected them. Section 4 contains a description of the methods used to investigate the two research questions and the different models fitted. In section 5, we first give a statistical description and summary of data then we analyzed the results relative to the research questions followed by a discussion on the findings. Finally, section 7 concludes this paper and gives some extensions of research that could follow this research.

## 2. Literature review on Performance Trends

Moore[4] has shown in his seminal paper about performance improvements in the Semiconductors published in 1965 [12] that performance (measured by the number of transistors on a microchip) is an exponential function of time, which later became known as Moore's law[5]. Since then, many scholars [13-16] have followed on studying technological progress and forecasting. In this report, to fit the performance trends and compute the improvement rate, we will refer to a generalization of Moore's law below (1) where $q_T$ is the relative performance of the hybrid variety (the yield ratio of the variety measured in bushel per acre) at year $T$ and $k$ represents the improvement rate:

$$q_t = q_0 e^{k(t-t_0)} \qquad (1)$$

More recently, *Benson and Magee (2015)* [17] and *Magee et al (2016)* [18], have shown that there exists an exponential relationship between performance and time for 28 different technological domains. More precisely, they authors [17-19] have quantified the performance improvement rate for 28 technological domains by examining empirical trends of performance using 71 metrics. These results show a wide range of improvement values (from 3-65% per year) for these 28 technological domain and are reported in Table1. Benson and Magee have also developed a method called the Classification Overlap Method (COM) [20], which provides a reliable, objective and automated method to classify patents data into technological domains where performance can be measured. *Benson and Magee* (2015) [17] found that information contained in patents (citations and patent publication year) are strongly correlated ($R^2 = 0.76$, $p\_value = 2.6 * 10^{-6}$) with the rate of technological performance improvement in that domain. In particular, they have established a statistically consistent model to predict the improvement rate of technological domains in which the coefficients of the predictive equation have been obtained by fitting 28 technological domain data. The predicted improvement rates for these 28 technological domains obtained were close to the observed rates. We will apply this predicting model on hybrid corn domain to obtain an estimation of the rate performance improvement in that domain and compare estimated and observed rates.

---

[4] Gordon E. Moore, the co-founder of Intel has shown in his 1965 paper an exponential growth in number of transistors in a dense integrated system which doubled each two years from 1959 to 1965 then fit forecast using his model.
[5] The label "Moore's Law" was given to this relationship by the Caltech professor Carver Mead.

| Technological domain | Rate K | Size of patent class | Relevancy |
|---|---|---|---|
| 3D-Printing (industrial stereo-lithography) | 37.60% | 251 | 93% |
| Aircraft Transport | 12.20% | 8629 | 79% |
| Camera Sensitivity | 15.60% | 1744 | 86% |
| Capacitor Energy Storage | 14.60% | 5944 | 84% |
| Combustion Engines | 5.70% | 19094 | 96% |
| Computed Tomography (CT) | 36.70% | 6817 | 88% |
| Electric Motors | 3.10% | 17869 | 86% |
| Electrical Energy Transmission | 14.90% | 10375 | 86% |
| Electrical Information Transmission | 14.30% | 44910 | 67% |
| Electrochemical Battery Energy Storage | 7.00% | 16122 | 83% |
| Electronic Computation | 33.00% | 13204 | 97% |
| Flywheel Energy Storage | 9.00% | 154 | 70% |
| Fuel Cell Energy Production | 14.40% | 7368 | 97% |
| Genome Sequencing | 29.30% | 3990 | 74% |
| Incandescent Artificial Illumination | 4.50% | 642 | 89% |
| Integrated Circuit Information Storage | 43.20% | 49018 | 81% |
| Integrated Circuit Processors | 36.30% | 149491 | 81% |
| LED Artificial Illumination | 36.20% | 3792 | 85% |
| Magnet Resonance Imaging (MRI) | 47.50% | 1778 | 86% |
| Magnetic Information Storage | 31.90% | 33576 | 93% |
| Milling Machines | 3.40% | 2315 | 93% |
| Optical Information Storage | 27.10% | 23543 | 82% |
| Optical Information Transmission | 65.10% | 36494 | 82% |
| Photolithography | 24.00% | 14975 | 87% |
| Solar Photovoltaic Energy Generation | 9.50% | 5203 | 85% |
| Superconductivity | 9.50% | 1776 | 85% |
| Wind Turbine Energy Generation | 9.20% | 2498 | 94% |
| Wireless Information Transmission | 50.40% | 39675 | 94% |

*Table 1: Improvement rates of 28 technological domains computed by Benson and Magee (2015)*

Triulzi and Magee (2016) [21] have developed[6] an alternative empirical model for predicting technological performance improvement rate. They have established several independent variables, which are the rate of technological obsolescence, the yearly rate of growth of impactful inventions, the centrality of cited patents, the search breadth (coefficient of variation of centrality of cited patents) and the competition factor and tested them across 8 different models. Their research reveals that the centrality of cited patents in the whole citation network comprising all USPTO patents, and the *yearly* growth rate of the number *of highly-cited patents* are the two variables that best estimate performance improvement rates in a technology domain. These two variables will be described more in detail in section 4 of the report and in section 5, we will compare the estimated rate of progress obtained from the two models cited above to the empirically observed rate of improvement in12 hybrid corn yield obtained from patent documents.

*Moser et al (2015)* [22] have recently published a NBER (National Bureau of Economic Research) working paper in which they analyzed the correlation between patent citations and size of inventive steps[7] from individual patented hybrid corn seeds where their data include 269 patents granted between August 26, 1986 and March 8, 2005 in subclass 800/320.1 *Maize*. They took the number of forward citations of patents as the dependent variable and tested an OLS (Ordinary Least Squares) and QML (Quasi-Maximum-Likelihood) Poisson regressions against independent variables such as improvement in yield, moisture and patent filing year as fixed effects. They have found a correlation between these variables and their regression results gave a $p\_value$ between $0.01$ and $0.05$ with $R^2 = 0.55$ using OLS models and $R^2 = 0.63$ for QML Poisson models. Following up on this work, we have also investigated such a micro point of view concerning individual hybrid corn varieties, to determine whether citation variables are correlated with the performance ratio due to specific seed varieties (similar to inventive step mentioned by Moser et al), through a study based on a sample of 70 hybrid corn patents granted between 1986 and 2015.

## 3. Data

3.1 Patent Database

We have set up a database that contains all patents of USPTO (United States Patent and Trademark Office) subclass 800/320.1 which corresponds to hybrid corn (also known as maize), that have been granted from January 1st 1985, the year in which firms started patenting hybrid corn seeds, to May 26, 2015 (when reporting the USPC in *Patentsview* database has been stopped). This database of 2,935 patents reports for each patent, the patent number, the hybrid variety name that has been automatically extracted, the name of assignee (the patenting firm or brand), the granted year and the number of US forward citations and claims.

The data have been collected from an USPTO website through the *Patentsview* tools which provides about 40 years of data on patenting activities in the United States including weekly updates. We first used a specific query based on *Patentsview* requests instructions [24] to download an html file that contained all information we needed (Patent number, Title, Firm, citations..) in html tags. Then pandas (a Python Data Analysis Library) has been used to extract information contained in the html file and save them into an excel file. And finally, we used a python code to read the excel file, clean the data (by deleting useless string values in the table values for example), extract important information as granted year and application year from the date format, then save the final output into a database containing for each patent, the number, the granted and filed year, the brand, the title, number of citations received and made by the patent and most important the name of the variety protected by the patent. To get the name of the variety, we first analyze the structure of the data and noticed that the name of the variety was contained in the title in most of the cases but after a deeper look, it happens that there were more than 50 different structures used for titles. We reported all these possible syntax of title and apply the code to extract only the variety name from the title for these possible syntax then checked up the output and made some manual corrections for some particular cases (see the code for more detail). An extract of the script used is in appendix. Figure 3 shows an extract of the initial raw data before being processed and the final clean database as output of the code.

---

[6] Their work (yet to be published), has been presented at ISS Conference at Montreal in July 2016, and is referred as 'Triulzi and Magee (2016), Predicting Technology Performance by Mining Patent Data.' For more information contact Giorgio Triulzi (gtriulzi@mit.edu).

[7] Moser et al use the term "inventive step" as the improvement in yield that occurs with a single patented hybrid corn seed. We will see that this is not easy to measure for hybrid corn but there is no other domain known where even a decent attempt can be made.

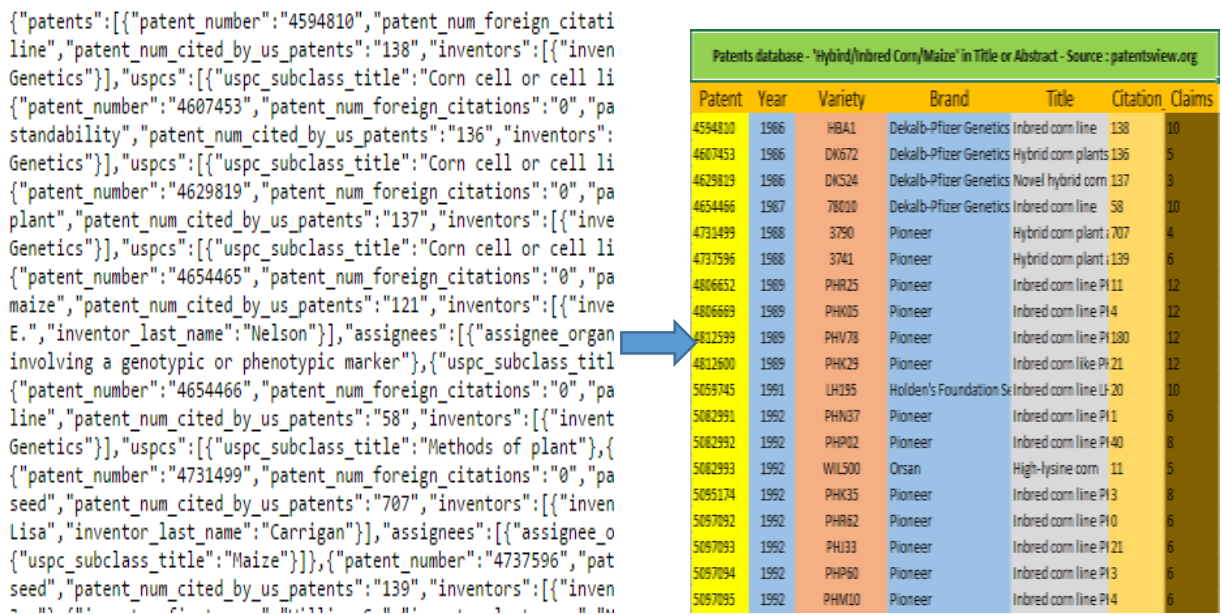

*Figure 3 : Extract of initial raw data on the left and final database at the right*

Patent documents on hybrid corn contain information relative to the patented variety, in particular the yield, moisture and other characteristics of the patented variety. This information is reported in tables within the patent and compared with others hybrid varieties values, 3 or more in general. The Yield (reported in bushel per acre) of the hybrid depends on the field-test, which means that the patented variety yield varies depending on the comparison variety and the test conditions. Table 3 illustrates an example on how the tests are reported, for a patent document where the patented variety has been tested against three different varieties. $P_i$ represents the value of the patented variety in the test $i$ and $C_i$ the value of the comparison, control variety C in the test number $i$. In the example below, A, B and C are different varieties tested in 3 different tests against the patented seed. The values reported in the patent document can refer to the yield, the moisture or other variables. Note that, a *variety X* is considered as better than *variety Y* in terms of Yield if *X* has a higher yield than *Y* while lower moisture is preferred. We noticed that most of patented varieties had a better yield than the comparison one but not necessarily lower moisture.

| Hybrid \ Trial | Test 1 | Test 2 | Test 3 |
|---|---|---|---|
| Patented Variety | $P_1$ | $P_2$ | $P_3$ |
| Control Varieties | $A_1$ | $B_2$ | $C_3$ |

*Table 2 : Illustration of trials values reported in patents documents*

We have constructed an excel file on hybrid corn patent data by reporting manually the yield and moisture values from patent PDF documents found on the Google patents database[8]. This database contains a sample of 70 patents of hybrid variety (not inbred) that have been filed from 1985 - year in which hybrid corn first became patentable- to 2014, including all hybrid corn patents filed until 1995 and at least two patents filed per year until 2010. By using this database, we computed a couple of variables that have been used to analyze the correlation with citations and also fit the performance trends of hybrid corn. The formulas are reported below where N is the number of tests made in one patent document, i.e. the number of comparison, each comparison correspond to different control varieties for a given patent, $P_i$ and $C_i$ are respectively the yield for the patented variety and control one in the test $i$.

- $Yield\_A = \frac{1}{N} \sum_1^N P_i$   -   The average yield of the patented variety among all tests
- $Yield\_B = \underset{i}{Max}(P_i)$ - The Best yield of the patented variety
- $Performance\_Ratio = \frac{1}{N} \sum_{i=1}^N \frac{P_i}{C_i}$   -   The average relative performance (yield) of the patented seed compared to the control varieties
- $Yield_M^Y = \underset{i}{Max}(Yield\_B_i)$ -   The maximum yield achieved for a given year $Y$ among all patented varieties filed in that year, which is equal to the maximum of Yield_B value of all varieties filed that year.

$Yield\_A$ values have been used to compare, for individual seeds, the average yield value reported in patent document to the average yield achieved by the same variety when tested in different states.

$Yield_M^Y$ will be computed for all years between 1985 and 2014 (patent filed year) to fit the performance trend and obtain the improvement rate in hybrid corn domain since it refers to the best value reported in the patent for a given year.

Figure 5 below shows a snapshot of excel files to illustrate how patented varieties data have been reported on excel files and the value averaged across different tests to get one average yield value for each patented seed. The table on the left side shows the structure for patent documents data (average taken across different comparison varieties) and on the right, the structure for the Illinois field-test database (where the average and standard deviation is taken across different regions)

---

[8] Through the link www.google.com/patents and search the number of the US patent

*Figure 4 : Snapshot of excel files containing patented seed yield from Patents and Field-tests*

Before going further, we need to clarify the difference between an inbred and hybrid variety since about 50% of the patents in USPTO subclass 800/320.1 (Maize) are actually inbred line instead of pure hybrid.

Self-pollination is defined as the deposition of the pollen grains from anther flower to the stigma of the same plant where in cross-pollination, the transfer is made to a different flower. An inbred line is obtained by self-pollination of the same plant over several generations. And due to the lack of genetic diversity, they lose vigor and became smaller and more uniform at each generation [39] but can provide interesting input for later hybridization. In fact, the hybrid variety is the result of a cross-pollination of two different inbred line (which are not self-pollinated). The hybrid became a larger plant than the two parents with better vigor thanks to the mix of the two different genes. Most hybrids are single-cross hybrid but there are also double-cross hybrids which result from a cross of two single-crossed hybrids (four parents) and also three way cross hybrids with three parents [40].

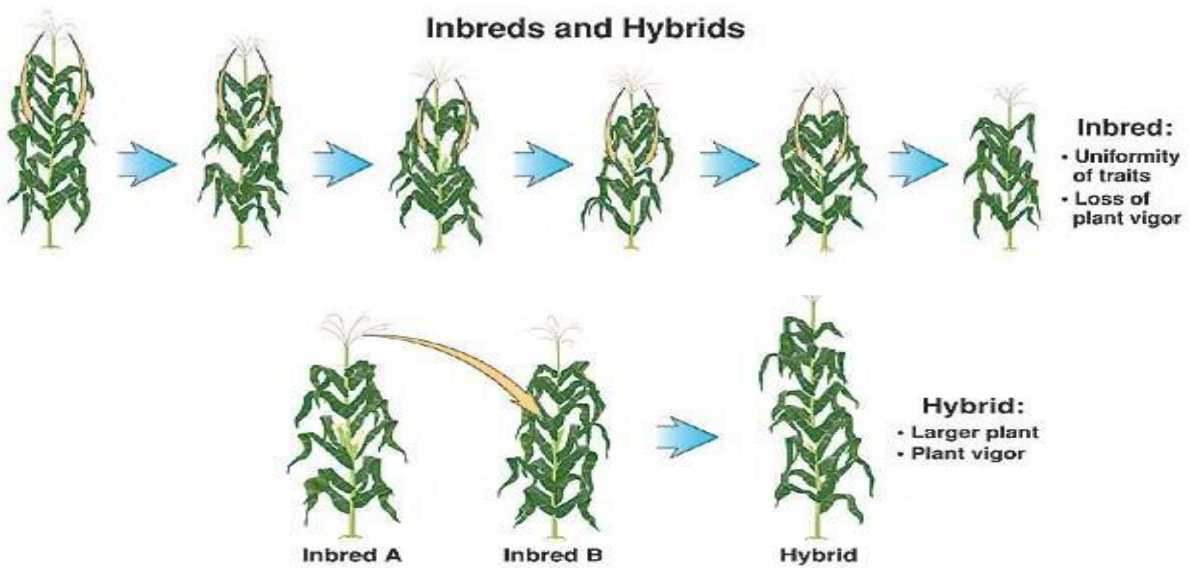

*Figure 5: Difference between Inbreds and Hybrids*

Now that we know the difference between these two types of varieties, it is important to note that *corn* and *maize* are synonyms and represent the same thing. In the following, when we mentioned about 'hybrid corn patents' we refer only to patents which contain the expression '*hybrid corn*' or '*hybrid Maize*' or '*maize variety*' or '*corn variety*' in the title and when we use the term 'Inbred patents', we refer to patents datasets having '*Inbred*' or '*line*' in the title of the patent.

### 3.2 States Field-test data

Land-Grant-Universities are institutions in the United States of America associated to some territories – States in general - that receive benefits from the Morrill Acts[9]. They were initially introduced to deliver higher education on agricultural science but also on engineering fields to embrace the industrial revolution: this system has evolved and most of them have become public state universities. Many of these Universities publish annual reports on field-testing of all hybrid corn varieties sold in their territories. These field-tests are an important data source for this research since our key purpose is to quantify the productivity improvement rate for hybrid corn and these data contain the yield in bushel/acre all varieties tested in any field in a given State. We will also use data reported in patent documents since they are an important indicator of innovation made in that domain and are the main input data to investigate on the micro-study. However, the State's field-trials involve much more data and thus remain essential. The State data also have limitations: the year coverage by these tests varies from one state to another and not all of them are accessible or downloadable. We chose to work with field-test data which cover a large year period to have a sufficient quantity of input data to better fit the performance trends.

Only 3 States which are Illinois (1995-2015), Ohio (1997-2015) and Kentucky (1998-2015) have both accessible and large amount of data then we gathered these information collected from University States websites. Ohio field-test annual data have been downloaded from the Ohio State University website extension [25].

We have set up a database containing Kentucky and Illinois field-test data, then use these information to compute the improvement rate in hybrid corn productivity but also some further statistical analysis on these trials. Kentucky and Illinois database analysis is detailed in the following subsection and Figure 10 shows a snapshot of these two states field-test files.

---

[9] Justin Smith Morrill first proposed Morrill Acts in 1857 and applied in 1862 which consist on granting federally controlled land to the states to be sell and raise funds that will used to establish and endow Land-Grant universities [35]

| Illinois Hybrid Corn Trials database - 1995-2010 - Source: http... |
|---|

| Year | Region | Brand | Hybrid | Yield | Moisture |
|---|---|---|---|---|---|
| 1995 | 1_Woodstock | AGRIPRO | AP453 | 152 | 19,2 |
| 1995 | 1_Woodstock | AGRIPRO | AP503 | 146 | 20,9 |
| 1995 | 1_Woodstock | AGRIPRO | AP9400 | 145 | 20,8 |
| 1995 | 1_Woodstock | AGRIPRO | EX9395 | 143 | 20,4 |
| 1995 | 1_Woodstock | ASGROW | RX484 | 138 | 17,6 |
| 1995 | 1_Woodstock | ASGROW | RX623T | 146 | 21,3 |
| 1995 | 1_Woodstock | ASGROW | RX707 | 153 | 21,8 |
| 1995 | 1_Woodstock | CALLAHAN | C7548 | 150 | 19,5 |
| 1995 | 1_Woodstock | CALLAHAN | C7645X | 145 | 20,3 |
| 1995 | 1_Woodstock | CALLAHAN | C7651X | 136 | 20,8 |
| 1995 | 1_Woodstock | CALLAHAN | C7654X | 146 | 22 |
| 1995 | 1_Woodstock | CALLAHAN | C7749X | 151 | 21,1 |
| 1995 | 1_Woodstock | CARGILL | 4127 | 146 | 18,3 |
| 1995 | 1_Woodstock | CARGILL | 4277 | 158 | 19,8 |

| Kentucky Hybrid Corn Tests database - 1998-2010 - Source: http://www2.ca.uky.edu/cornvarietytest/Archives.html |
|---|

| MATURITY | YEAR | BRAND | HYBRID | YIELD | MOIST | STAND |
|---|---|---|---|---|---|---|
| EARLY | 1998 | PIONEER | 33G26 | 190.3* | 14.4 | 96.2 |
| EARLY | 1998 | PIONEER | 33V08(Bt) | 186.3* | 13.8 | 100.0 |
| EARLY | 1998 | CAMPBELL SEED | 7250 | 182.9* | 14.2 | 99.1 |
| EARLY | 1998 | AGVENTURE | A.V.845 | 178.5 | 14.4 | 97.5 |
| EARLY | 1998 | PIONEER | 33Y18 | 178.0 | 14.5 | 99.3 |
| EARLY | 1998 | DEKALB | DK626BTX | 177.3 | 13.4 | 99.6 |
| EARLY | 1998 | BECK'S | 5912W | 177.0 | 15.3 | 100.0 |
| EARLY | 1998 | BECK'S | 6205 | 176.8 | 14.3 | 99.9 |
| EARLY | 1998 | DEKALB | DK635 | 175.7 | 13.6 | 99.0 |
| EARLY | 1998 | CALLAHAN | C7557 | 174.8 | 14.2 | 99.3 |
| EARLY | 1998 | CROW'S | 496 | 174.5 | 14.3 | 99.6 |
| EARLY | 1998 | GARST | 8464 | 173.7 | 14.5 | 99.5 |
| EARLY | 1998 | DEKALB | DK618BTX | 173.5 | 14.1 | 98.5 |
| EARLY | 1998 | MYCOGEN | 2725 | 172.9 | 14.1 | 100.0 |

*Figure 6 : Snapshot of Kentucky and Illinois field-test database*

3.2.1 Kentucky

Kentucky Hybrid corn trials reports were available only on PDF documents [26], so we first copy-pasted them into an excel file, then use excel tools and some manual process to correct mistakes that occurred after the automation. For example, many cells were shifted to the wrong columns on each line where there was irregular space on the PDF document or when the variety brand name was in two words (LG SEEDS, NK Brand, Seed Consultants, Garst Seeds, Croplan Generics, Steyer Seeds, Great Lakes…). The annual reports gave yield and moisture values for different counties and regions but also an average value for all regions (in contrast to Illinois field-tests which do not include average across regions). We keep track of the latter since we needed one productivity value for each variety.

To get the patented variety among the tested varieties, we run a Python code with the patent database and the Kentucky database as inputs data. We found that from 1998 to 2010, 28 varieties were patented. We study in particular 14 patented varieties (13 hybrids and 1 inbred) from 1998 to 2005 compare their performance to the average performance on the test and we found three interesting points to note. First, the patented varieties are all among best performing one and have a better yield than the average yield in that test and their moisture[10] was close to the average moisture in the test as we show on Figure 7 (yield is in bushels/acre and moisture is percentage, each variety is represented by a number on the *x-axis*). Secondly, the firms filed the patent application for a variety before or in the same year that the first test of that variety, which suggests that the firms are aware of the variety performance before field-test were performed, possibly because they do their own tests before. Thirdly, we found that the maturity of the hybrid doesn't impact on the seed performance as shown on Figure 8 where the reported yield is almost the same for the 3 maturities level.

---

[10] A lower Moisture is preferred while a higher Yield is better

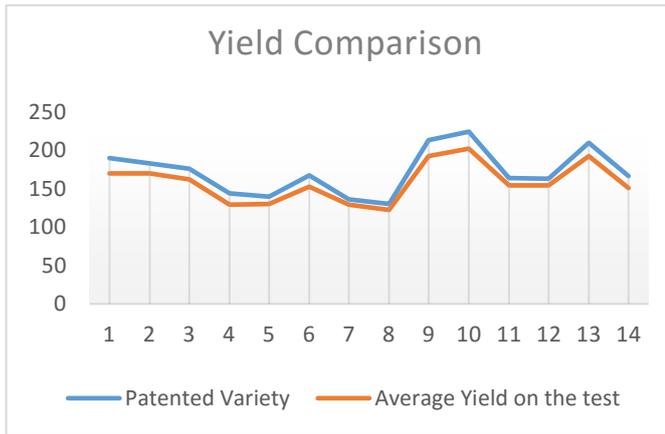
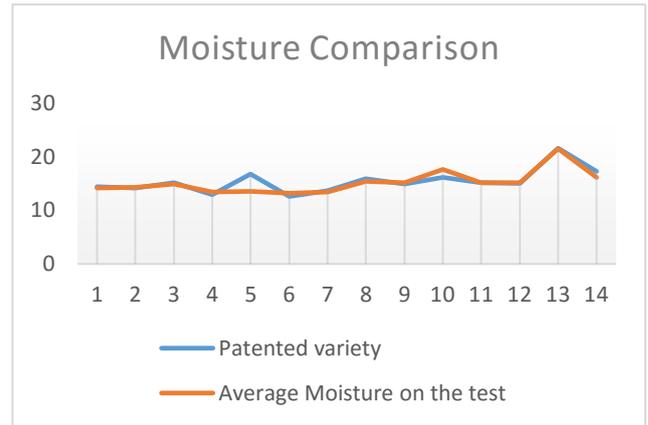

*Figure 7: Kentucky patented varieties Yield and Moisture*

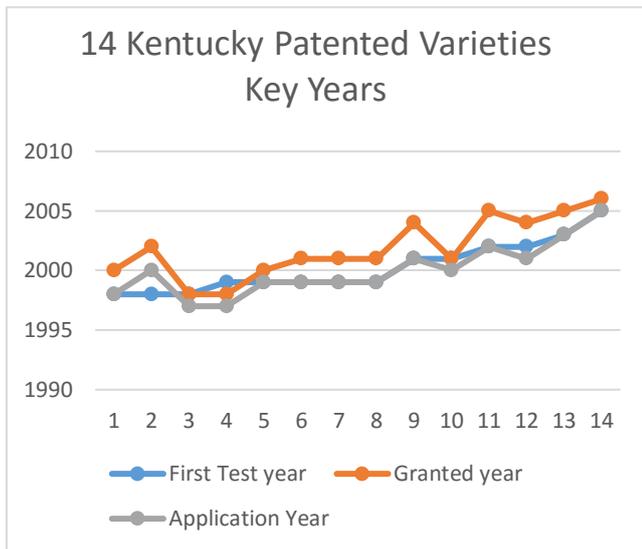
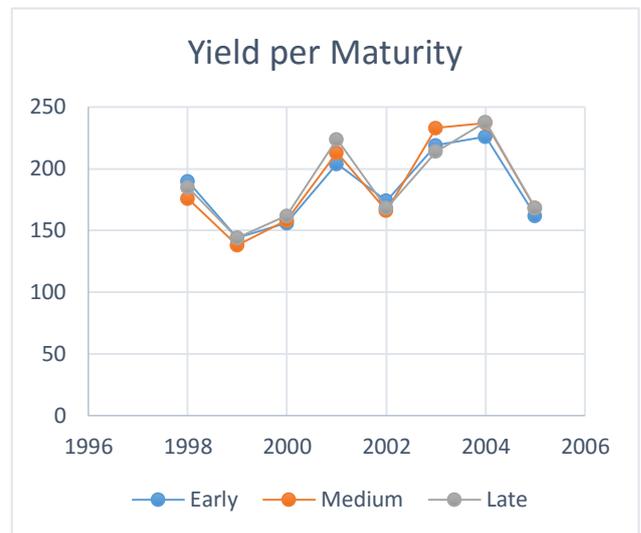

*Figure 8: Maturity level and patent year of Kentucky patented varieties*

3.2.2 Illinois

We have also studied Illinois field-tests on hybrid corn and undertaken a comparison between the yield value reported in the patent document and the one reported in the state trials. Contrary to Kentucky, Illinois does not provide average value across different regions, so we have downloaded excel tables per region for each year on their website [27] and merged them on one excel file.

A first statistical analysis of these tests showed that in total from 1995 to 2015, 5,434 different varieties have been tested and the number of tested varieties and the seed brands has considerably decreased, from 680 varieties and 78 firms in 1995 to 222 varieties and 29 firms in 2015 (Figure 9). This drop in number of firms (and also varieties) is in part due to the firms' consolidation introduced above. In particular, we investigated on the patented seeds in the field-trials from 1995 to 2005 - that have been obtained using a python code and we found that only 29 varieties have been patented ( and we got 41 if we include the duplicates tested over other regions). After 2005, we have found only one more patented seed but the number after 2005 is not exhaustive because it doesn't take into account Monsanto varieties. Indeed,

Monsanto (a firm) which bought Dekalb (originally a firm but now a brand) in 1998 keeps using the Dekalb name in seeds they sell. In Illinois tests, the seed names are designated by two letters (DK) followed by four digits. While, in the seeds they patent, another uniform nomenclature is used for Monsanto patents, and the patented seed name is written with two letters (CH) followed by six digits. Which means that, we couldn't link Monsanto tested seeds in Illinois *(DKxxxx)* to Monsanto patented seeds *(CHxxxxxx)*. Independently on Monsanto varieties, having only 30 Pioneer seeds which are patented among field-test starting 1995 is a relatively small number, but if we investigated deeper on field-test data and firms who own these seeds, it tends to became more clear.

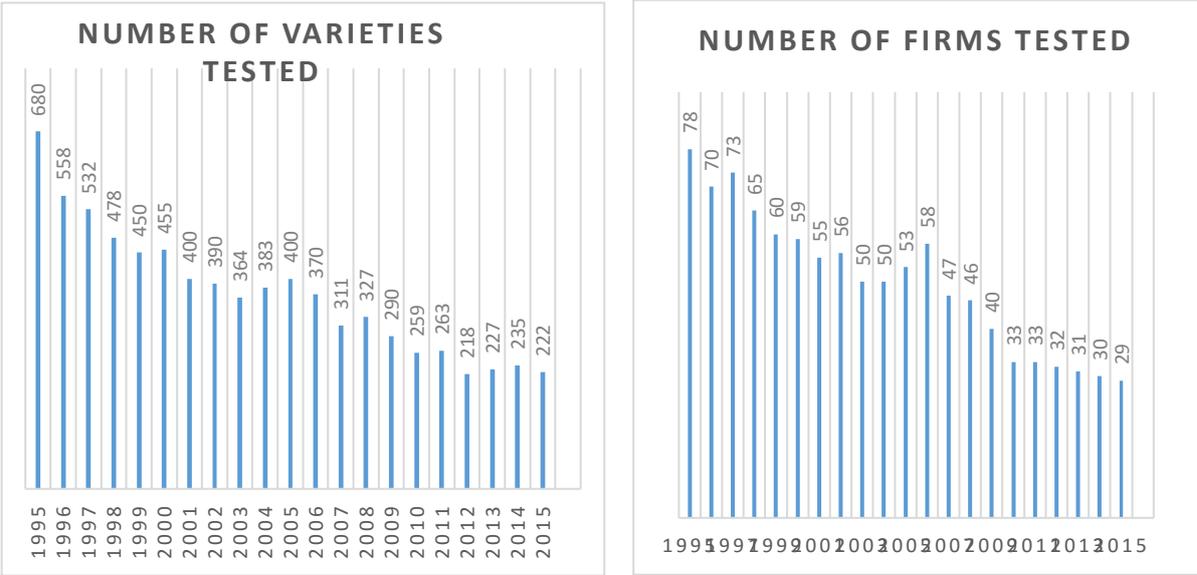

*Figure 9 : Number of varieties and firms tested I Illinois field-tests 1996-2015*

Indeed, these small number of patented varieties makes sense when we examine in detail the repartition of firms and brands that own the seeds patented (2 935) and tested in Illinois (5 433). As we can see on Figure 10, for the same firm their share on the seeds patented (left) and tested on the field-tests (right), Pioneer and Monsanto who own 75% of patented seeds, own about 50% of tested varieties (20% from Pioneer in addition to 28% Dekalb-controlled by Monsanto- and other subdivision of these two). In particular, Golden harvest and Garst hold more 35% of tested seeds in the state trials where they own less than 1% of patented seeds. We can say that the fact that only few firms are patenting seeds while a larger number of firms seeds are tested in the field-test justify in part this small number of patented varieties found. Moreover, the relatively small number of Pioneer seeds patented (30) from 1995 can be explained by the small percentage of Pioneer seed tests (from 23% to 5%). Indeed, from 1995 to 2005, 96 Pioneer seeds have been in field-tests, which represent about 23% of the number of Pioneer granted patents (415) for the period, while from 2006 to 2015, only 68 Pioneer seeds have been tested, around 5% of the number of Pioneer patents granted in the same period (1 190).

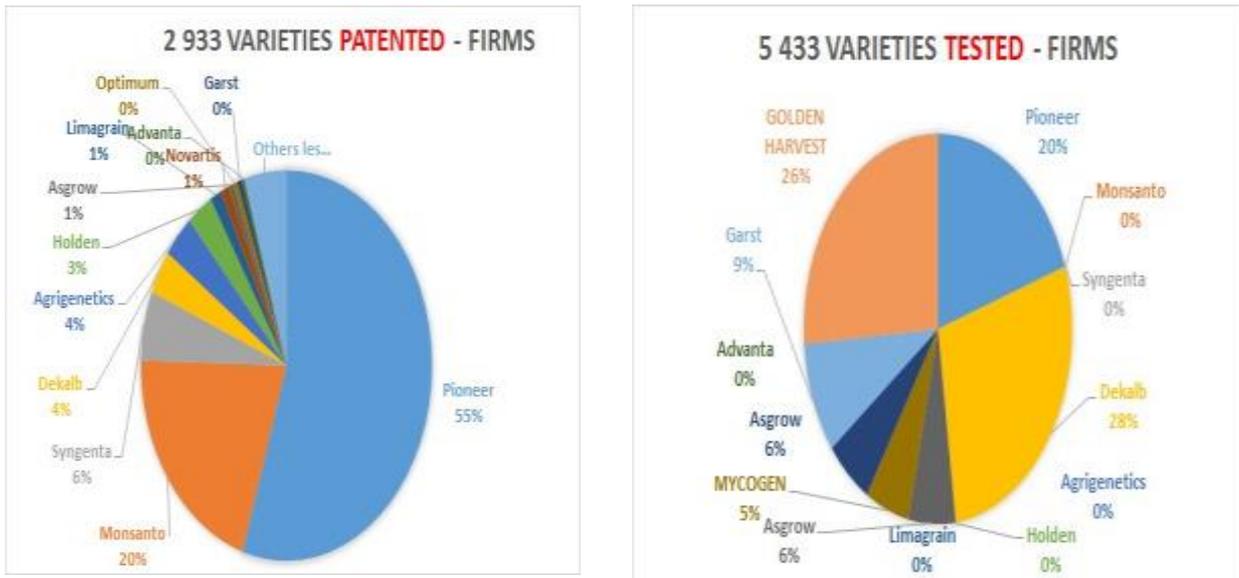

*Figure 10: Firm shares on Patented seeds and Illinois tested seeds*

We worked on the 30 patented varieties tested in Illinois trials to compare the yield achieved by the individual seeds in the field-test to the yield reported in patent documents, i.e we investigated whether there is a high difference or not in the yields reported in two different data sources. In Figure 11, we compare the absolute yield from patent and state data and note that there is no particular dominance, since depending on the variety the yield from the patent document can be higher or lower than what is reported in state data (see left graph). On the right graph of Figure 11, we analyzed the relative yield of the same 30 varieties by comparing 3 performance ratios: the yield of the patented seed over the controls varieties from the patent document (YLD/Comparison in blue), the yield of the patented variety over the best yield in the field-test from state data (YLD/Best in red) and the ratio of the patented seed over the average yield of variety tested in the field-test (YLD/AVG in gray).

Note that, for the varieties tested in more than one year in the same region, we referred to the yield values achieved by the variety in the first year in which the variety was tested to calculate the average and best yield values as. The absolute and relative yield both indicate that the yield reported in patents is close to the average yield of varieties tested in a region in field-tests, which gives one confidence that patent trials and field trials are done in common conditions.

For each of these 30 varieties (represented in the x axis) we have also compared the absolute yield of the patented variety to other varieties tested in the field-test. As shown on Figure 12, we noted that the yield of the patented variety (in blue) is very close to the average yield achieved in the field-test (average for a given region where the variety was tested) and any of these patented varieties are the best performing ones since their yield is lower than the best performing seeds yield (in orange). This differs from the Kentucky field-test results where the patented variety in the field-test was always higher than the average yield and often close to the best performing one. We have also observed a high standard deviation - due to the weather and soil particularities - on average yield values of the tested variety average across different regions where the variety have been tested. While the average values from patent data have a much lower variation probably because the tests made by patents applicants are done in controlled conditions.

We now have an overview on Kentucky and Illinois field-test data including a comparison with patent data for Illinois trials. We will describe the method and findings in the two next sections to attempt to respond to the two research questions described in the introduction.

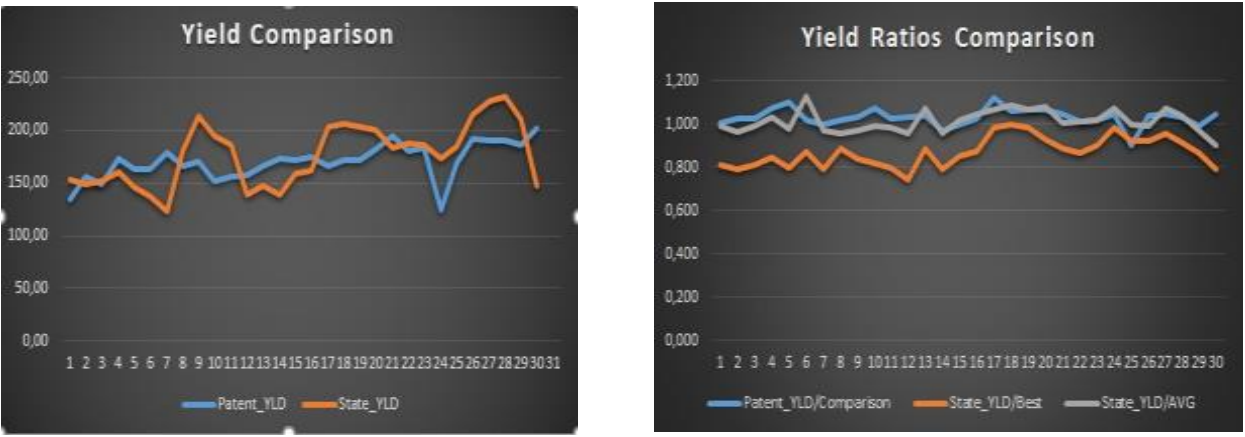

*Figure 11 : Absolute and relative Yield comparison from Illinois Trials*

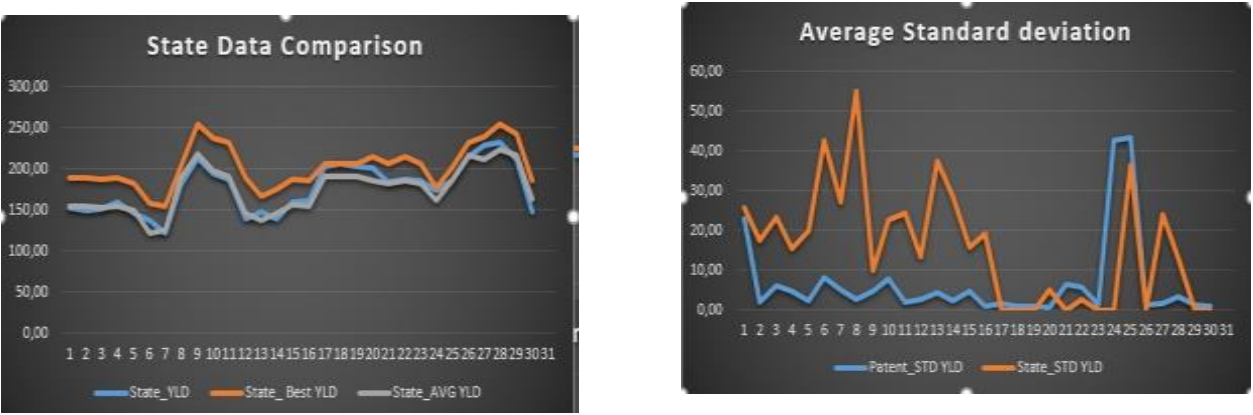

*Figure 12: Illinois Patented varieties comparison to best performing ones*

## 4. Method

This section develops the different models and methodology that will be used to attempt to answer the two research questions of this paper. The first research question was, for a macroscopic analysis, to calculate the productivity growth rate on hybrid corn domain using Moore's law by fitting productivity trends using both patent data and 3 States (Ohio, Kentucky and Illinois) annual field-test data. The second research question is - for a microscopic analysis, at the individual seed level - to find out whether the best performing varieties receive more citations than other varieties.

4.1 Performance improvement rate

4.1.1 Empirical rates

The main purpose of this research is to calculate the observed productivity improvement rate on hybrid corn domain using empirical data. For that, we analyzed performance trends from several sources and for different year period. This analysis can be break down into three parts, three sets of trends.

The first set of trends are fitted using empirical productivity data (yield in bushel per acre) from Illinois's, Kentucky's, Ohio's field tests and from the patent documents, for the same year period between 1996 and 2015 (the maximum coverage we could get for these 3 states). Note that, Ohio State University provides annual productivity data on hybrid corn starting 1997. To get the average yield data achieved in 1996, we used a simple mathematical relation by inferring 1996 average productivity from 1997 annual data and two-year summary (i.e. covering the period 1996-1997) average data which was also provided. We used the same method to get Kentucky average productivity in 1997 (using two-year summary data from 1998) and in 1996 (using two and three years' summary data from 1998).

The annual performance data used for this first set of trends are all average data, i.e they represent the average yield achieved in field tests by different hybrid corn varieties. For the patents, we took the average yield of the patented variety reported in the patent document patents ($Yield\_A$ variable) and for the 3 states, we took for each year the average value of yield achieved in the field-test across different regions (or county) tested. For example, the productivity value taken for Illinois in 2000 is actually the average of the 4 regions (North, South, East central and West central) average yield. We note that these values from states data are not only subject to weather variations but also to the regions soil properties including probably a high standard deviation (for example before 1998, Illinois trials were reported by county, from 7 to 10 different counties). We will also fit the trend using calculated U.S. average (across States) corn productivity data from USDA (United States Department of Agriculture) [28] for the same period covering the last twenty years and compare these trends and the resulting improvement rate.

The second set of trends uses data sources come from patents only and for different time period. These trends have the objective of being more precise than the first set by using the best yield (not average) value reported in patents ($Yield\_B$ variable) applied for in one year. The best yield achieved by hybrid corn varieties that seek patent protection in a given year is more likely to quantify the progress made by hybrid corn technology from year to year, compared to average values. However, a problem with this method lays in the fact that our dataset includes only 4 patents after 2010, (because most of firms stopped reporting the yield in bushel per acre but in another metric). This makes our patent dataset less representative of what innovations have been made in last 5 years. To assess how serious this problem is, we fitted two performance trends from 1985 to 2010 and from 1985 to 2015, using the best yield value reported in patents documents for each year ($Yield_M^y$ variable).

Finally, the third type of trends fitted aims is to provide more accurate estimations of the improvement rates by correcting some limitations of the first set of trends fitted. For that, we do not take average value per region or per year as input data but the entire Illinois field-test database which contains more than ten thousand varieties productivity data from 1995 to 2010. Indeed, the States' data used to compute the improvement rate in the first part are subject to weather changes and variations in soil across regions (Figure 11 shows large data standard deviation) since they represent average values across different regions. To tackle this issue and get an improvement rate cleaned from weather changes and see how weather changes impact on the estimated K from states average, we used a method similar to a 'differences-in-differences' analysis. First we searched for varieties that have been tested in several consecutive years in the same region. Variations in yields over time of the same variety planted in the same region are likely to be largely due to weather variations. Therefore, their yield can be used as control variables. We then calculated the highest yield achieved by a variety tested in the same region in the same year. Finally, for each year, we divided the yield of the best performing variety by the yield of the control variety, and then fit the performance trend. This trend can be explained by the performance trend of the best performing variety for hybrid corn, correcting for differences in soil between regions and variations in weather between and within regions and across time. This method will be applied on Illinois field-tests and the resulting improvement rate of that trend will be compared to the one resulting from parts 1 and 2 described above in section 5 and see whether the weather variations has an important effect on the yield achieved and the resulting improvement rate.

4.1.2 Estimating models

In this subsection, we will first explain the predicting models, then note how the empirical rate is computed using patents and states averaged data and finally explain an alternative method to compute the improvement rate corrected from weather variations impacting yields reported in field trials performed in different States.

The first estimating model, as introduced in the section 2, has been developed by Benson and Magee (2015) [17] uses patent data to estimate technologies performance improvement rates. This predicting model gives an improvement rate $K_1$ using 2 independent variables which are the average publication year of patent $AvePubYear$ (which measures the recency of a technological domain) and the average number of forward citations received by a patent within 3 years ($Cite3$) after the publication year. These two metrics are calculated by computing equation (2) and (3), where SPC is the simple patent count (number of patents), $FC_i$ is the number of Forward citations for patent $i$, $t_{i_{pub}}$ is the publication year of patent $i$, $t_{ij_{pub}}$ is the publication year of the forward citation j of patent $i$, and function $IF(arg)$ only counts the values if the argument is satisfied.

$$Cite3 = \sum_{i=1}^{SPC} \sum_{j=1}^{SPC} IF(t_{ij_{pub}} - t_{i_{pub}} \leq 3) \quad (2)$$

$$AvePubYear = \frac{1}{SPC} \sum_{i=1}^{SPC} t_{i_{pub}} \quad (3)$$

By computing these variables[11] for a given domain, we can now calculate a time-based technological improvement rate K1 with the equation 4 (model 1) where the coefficients have been estimated using data on 28 technology domains by Benson and Magee (2016).

$$K_1 = -31.1285 + 0.0155 * AvePubYear + 0.1406 * CiteF3 \quad (4)$$

The second empirical model developed by *Triulzi and Magee (2016)* [21] has also two independent variables, the centrality of the patent and the rate of growth of impactful inventions (called 'Z') which are calculated using the equation (5) and (6). First, the authors computed a measure of the centrality of a patent in the whole citation network comprising all patents granted by the USPTO. The centrality is computed using a measure called Search Path Node Pairs (SPNP) (reference). The SPNP value for each patent *i* is computed by multiplying the number of paths connecting patent *i* to any other patent granted before, for the number of paths connecting patent *i* to any other patents granted after *i*. SPNP is essentially a measure of information flow centrality. If we take a random walk from any patent to any other patent following citation paths, it is more likely to come across patent *i*, if its SPNP value is large. To remove any confounding effect due to the growing size of the citation network, the SPNP value of each patent *i* is normalized as a rank percentile of the distribution of SPNP values for patents with the same application year of patent *i*. compared to all other patents by taking the rank of the patent In that equation. This is what we define as $RankPerc_i$. $CB_i$ is the number of patents cited by patent I and SPC is the total number of patent in the domain of interest.

---

[11] Note that, in this research, we computed *Cite3* and *CiteForward* (number of citations received in total) variables for patents 70 patents granted between 1986 and 2015 excluding 4 highly cited US patents (4629819, 4607453, 4731499, 4737596) and 6 patents filed after 2010.

$$Centrality = \frac{1}{SPC}\sum_{i=1}^{SPC}\frac{1}{CB_i}\sum_{j=1}^{CB_i} RankPerc_j \qquad (5)$$

The rate of growth of impactful inventions (which the authors call 'Z') represents the slope of the exponential relationship between Year and *CumSum* where *CumSum* is a vector that contains the cumulative sum of number of highly cited patent per application year contained in the vector Year. Z is calculated using Equation (6) where SLOPE (X, Y) is a function that returns the slope of the relationship of Y over X and Log the natural logarithm function that can be applied to a vector.

$$Z = SLOPE\ (Year, Log(CumSum)) \qquad (6)$$

The rate of performance growth $K_2$ can now be computed using the equation (7) :

$$K_2 = \exp(5.0575 * Centrality + 10.1261 * Z - 5.8486) \qquad (7)$$

The authors of the second estimating model have computed the equation above to calculate the yearly improvement rate in 28 technological domains and their finding is illustrated below in the Figure 13 (exponential of observed K against exponential of estimated rates K). As we can see it, the results are confident with a Pearson coefficient of 0.86 which indicates that the predictions are reliable.

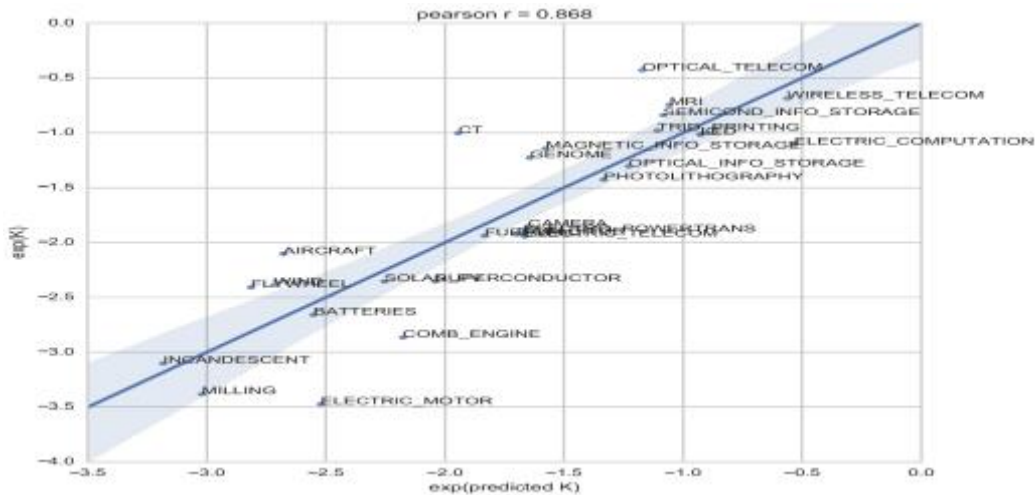

*Figure 13: Observed improvement rates over estimated ones by Triulzi and Magee (2016)*

For this paper, K1 will refer to the rates estimated by Equation (4) and K2 the one estimated using Equation (7). In section 5, we will compare K1, K2 and empirical rates of growth obtained from patents and states data. In particular, because of the changes observed in patenting practices, we will compute these rates using patent data (to compute variables) from two different datasets: a dataset with patents whose application year is until 2005 (before the patenting practice change occurred) and patent data up to 2013 (since the initial model in equation (4) has been developed using patent data until July 1st, 2013 for 28 technological domains).

Since there is an important difference between hybrid and inbred varieties, not only from their seeds development but also in the repartition of number of patents applications, which is likely to impact on the predicting models' parameters. Consequently, to be more precise, we will estimate rates separately using only inbred varieties, hybrid varieties and both of them, form patents that has been filed until 2005 and until 2013 (6 datasets).

## 4.2 Regressions models on citations

We hypothesize that high performing patented seeds will receive more follow on inventive work and, therefore, receive more citations from other patents. The second micro-aim of this research is to investigate the correlation between citations received by a patent and the relative performance of the patented variety. For that, we tested several models across variants of the variables where the dependent variable is always a variable of citations received by a patent and the independent variables are the performance ratio of the variety (which equation is noted below), the publication year of the patented variety or the absolute yield value.

The models tested are summarized in Table 3 where the main independent variable is the relative performance of the patented variety over the control variety defined as follow where $P_i$ represents the yield in bushel per acre of the patented seed in the test I and $C_i$ the yield in bushel per acre of the control variety in the test number $i$ and N the total number of comparison tests made in the patent document.

$$Performance\_ratio = \frac{1}{N} \sum_{i=1}^{N} \frac{P_i}{C_i}$$

The average number of forward citations within 3 years (Cite3) is a good potential indicator of immediate importance of a patent and the rank percentile of citations which correct the citations number from year effects is also an interesting variable to analyze. Thus, model 1, 2 and 3 are of interest because first, they will allow us to compare the correlation between these 3 types of citations (the total forward citations received by a patent, Cite3 and Cite3 rank percentile[12]) and see which one is the most correlated to performance ratio as determined relative to individual seeds and patents. Secondly, model 1 will indicate whether there is a correlation between total forward citations received by a patent and the performance of that variety. Furthermore, model 1 results will be compared to Moser et al [22] results since the same model has been tested in their published working paper and they found an effect.

| Model\Variable | Dependent | Independent |
|---|---|---|
| 1 | $CiteForward$ | $Performance\_Ratio + Year$ |
| 2 | $Cite3$ | $Performance\_Ratio + Year$ |
| 3 | $Cite3RankPercentile$ | $Performance\_Ratio + Year$ |
| 4 | $CiteForward$ | $Performance\_Ratio$ |

*Table 3 : Models tested with OLS and Poisson regression*

The model 4 will be run to investigate whether without the year fixed effects in model 1 (independent variable Year), we have a correlation between the total citations and the performance ratio or not.

---

[12] Note that Cite3RankPercentile contrary to Cite3 and CiteForward variables is not a count data and is bounded by 0 and 1, thus they best model to analyze for this dependent variable effect is logistic regression, consequently, we do not expect a strong correlation with this variable by using OLS or Poisson Regressions.

Since citations are count data and a Poisson regression is known to be more adapted to count data, Poisson regression will be computed for the first 4 models. An OLS (Ordinary Least Squares) regression will also be run for these 4 models as a secondary source of comparison even though we expect Poisson regression results to be more significant than the simple linear model. The 4 models listed above will be tested using patent data filed between 1985 and 2014, excluding 4 highly cited patents and 4 other patents that have been recently filed. Moreover, since we have noted a drastic change in Hybrid Corn patenting practices, we will run a Negative binomial model on patent data filed until 2008 for the four models and analyze whether the result obtained with Poisson Regression are confirmed by the Negative Binomial model or close. Negative Binomial regression tends also to be more adapted to our data since we observed have a high standard deviation value compared to the mean of the performance ratio.

# 5. Results

This section gives an analysis of the results obtained regarding the two research questions announced, respectively in subsections 5.2 and 5.3. But before that, we will give some summary results on patents data we used, in particular, we will give a detailed analysis on hybrid corn patenting firms and seed industry in the main, including the important consolidation system build over years which led to a very few number of firms that controls patenting seeds. We will also use the patent data gathered to investigate on patenting practices and citations in patent evolution over year.

5.1 Hybrid corn industry and patenting changes

      5.1.1 Seed industry

In this section, we will first describe the identification of firms and brands in our patent dataset, then give an overview on the seed industry in general and its consolidation. We noticed that only a few firms hold the majority of patents over the last 30 years. In particular, Pioneer[13] and Monsanto hold 75% of the patents in the hybrid corn domain where Pioneer alone holds more than 50% of patents and the other 25% are shared between small firms, as shown in figure 14 below. We note that the Dekalb brand, as many others, was purchased by Monsanto in 1998 and the chart in figure 14 show the overall consolidation for all years combined. It is interesting to look at the evolution of the number of patenting firms for the last three decades as shown in figure 15 which also includes at the top left the number of patents per year period. This chart reveals that about 50% of our patents dataset (more than 1 400 of them) have been granted these last 5 years which means that the less firms are patenting many more seed varieties.

---

[13] Pioneer, formerly Pioneer Hi-Bred is now labelled as Dupont Pioneer after they merge with the American conglomerate DuPont in 1999, is one of the mains U.S. producers of Genetically Modified Organisms, including genetically modified crops with herbicide and pesticide resistant

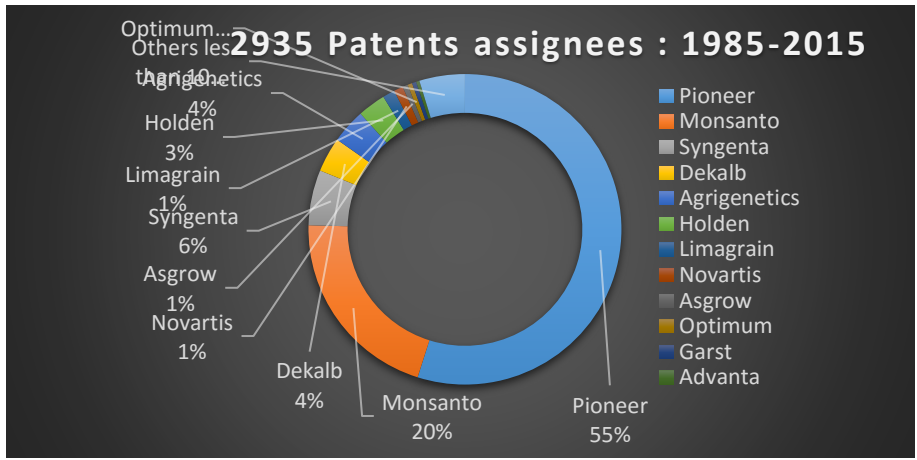

*Figure 14 : Patenting firms shares in hybrid corn varieties*

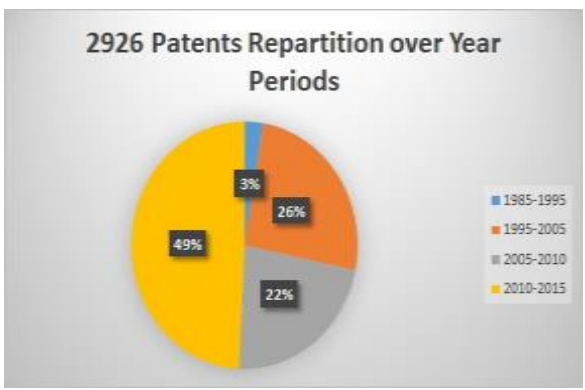
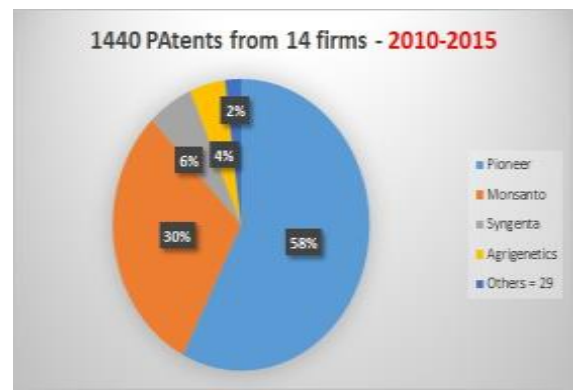

*Figure 15: Patenting Firms repartition over four years periods*

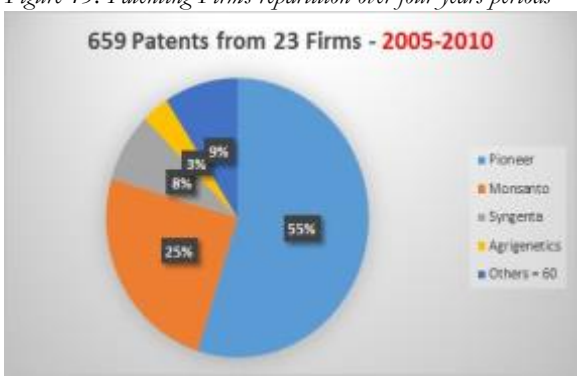
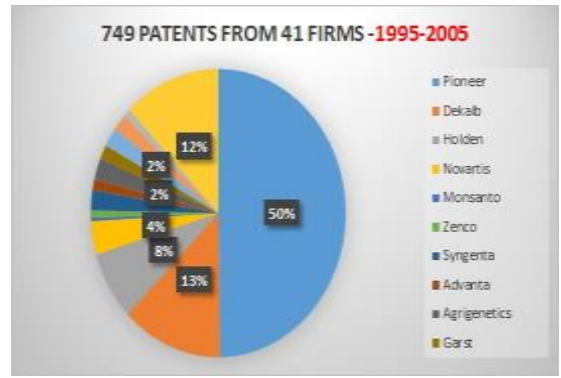

This oligopoly observed in hybrid corn patents is generalized even in the overall seed industry. Indeed, according to the ETC Group [30] a Non GMO organism that criticizes consolidation in the seed market, in 2013 60% of the market is controlled by only three firms: Monsanto, DuPont Pioneer and Syngenta. As shown on figure 9, Monsanto and DuPont Pioneer controlled in 2014 about 70% of the corn seed market and more than 60% of the soybean seed market. This industry consolidation, through purchases and mergers has accelerated since the early 1900s [30] with a few firms such as Dupont, DowAgroscience and especially Monsanto in 2000s starting to purchase many smaller seeds companies.

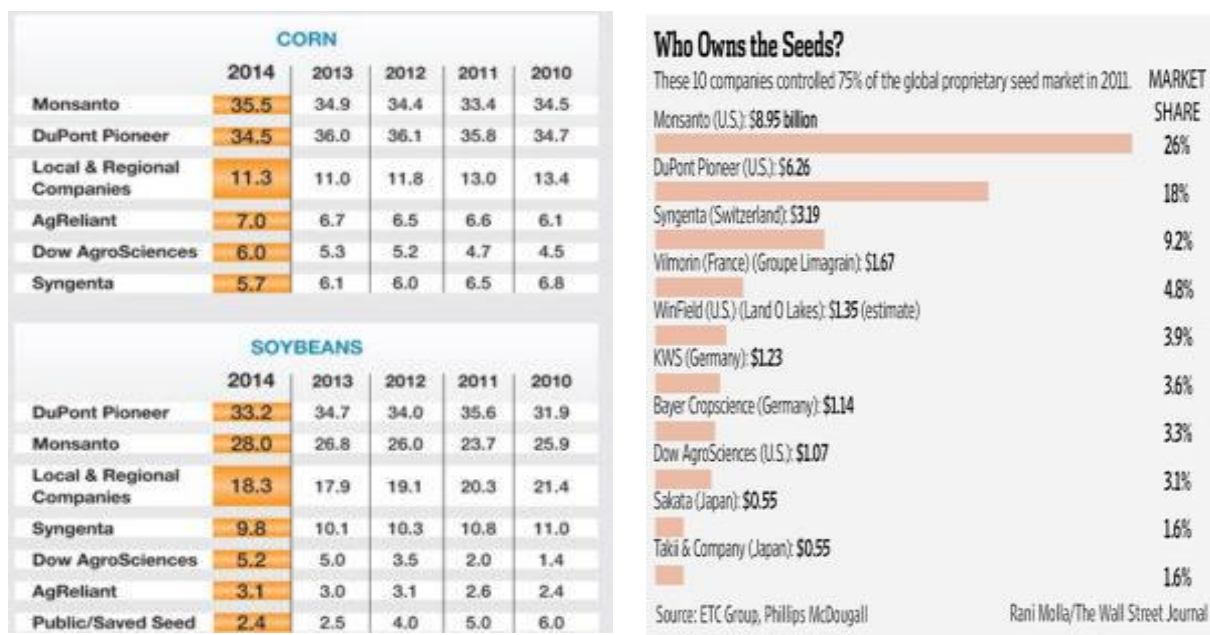

*Table 4: Seed market shares repartition*

For example, as illustrated in the seed family tree [31], DuPont completely purchased Pioneer in 1999 after a first acquisition of its 20% shares in 1997 and Syngenta is the result of a merge between Novartis and AstraZeneca in 2000. The most recent important merge announced in 2015 [32], is between Dow and DuPont[14] (which owns Pioneer) becoming DowDupont Agri which will lead the group in the second position of the global seed market (21%) behind Monsanto (26%) and in second position as well in the Pesticides market (16%) behind Syngenta (23%). The consolidation has been criticized by anti-GMO activists who accused them of taking control of world's food supply reducing farmers' options. Vandana Shiva, a philosopher and anti-GMO has denounced the seed industry business and said that [30] : " *Seed slavery is ethically important to address because it transforms the Earth family into corporate property. It is ecologically important because with seeds in the hands of five corporations, biodiversity disappears, and is replaced by monocultures of GMOs. … In our times some corporations think it is alright to own life on earth through patents and intellectual property rights (IPR). Patents are granted for inventions, and life is not an invention. These IPR monopolies on seeds are also creating a new bondage and dependency for farmers who are getting trapped in debt to pay royalties.*"

This consolidation system tends to indicates that patenting seeds is highly controversial but it has become a business tool in this vast seed industry which makes more complex hybrid corn patent analysis and seems to have resulted in some important patenting practice changes.

---

[14] DuPont is the world's fourth largest chemical company based on market capitalization and eight based on revenue while Dow is the third largest chemical company by revenue in 2015 (after Sinopec and BASF) [36].

5.1.2 Patenting changes

There is clear evidence of a drastic change in patent practices in the hybrid corn domain starting from 2008. Figure 16 shows that the number of patents per year in the total USPTO and in inbred patents do not change noticeably from 2008 but that the number of hybrid corn patents increases by a factor of 4 roughly (which increased about 50 before 2008 to 200 after 2008). For example, the number of hybrid corn patents filed per year was always below 100 before 2008 then from that period to 2010 more than 600 patents have been filed. Following on these patenting practices, we wondered whether a notable change has occurred in citations practices and the findings indicate so.

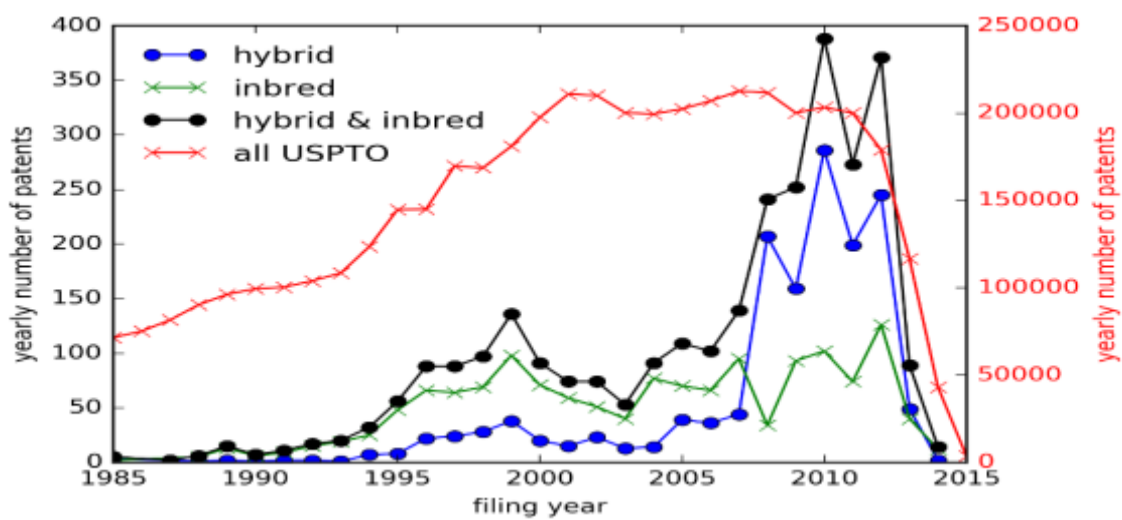

*Figure 16: Number of patents applications per domain*

Indeed, when we examined the evolution of backward citations (citations by Hybrid Corn patent to other patents) over year, we noticed two interesting points. First, the average number of backward citations (citations from hybrid corn patents to others patents) has considerably increased (more than quadrupled) after 1998 since that average number jumped from 2 citations per patent in average before 1998 to 9 citations in average per patent after 2003 as shown on Figure 17 where the x-axis refers to patent filing year. However, note that these mean values are not very representative of individual patents number of citations, since they have a large spread, especially during the last years where we have observed a very high standard deviation. As show on the right graph of Figure 17, the standard deviation (in orange) is sometimes even higher than the average number of citations (ABC in blue). Regardless of the spread of the number of citations, this drastic rise in backward citations will certainly impact on the Cite3 variable and eventuality on the Centrality variable as well which are the key parameters of the estimating models used. Thus, we expect our estimated rates obtained from these models to be less confident and reliable compared to the empirical observations. Another point noticed about the citations analysis is that the number of citations made by hybrid corn patent to others applications (only filed, not granted yet) has also know an important rise after 2008. In particular before 2008, the average number per year of citations to applications was around 1 while after 2008, that number of citations has risen to 3 and 4 in average per year.

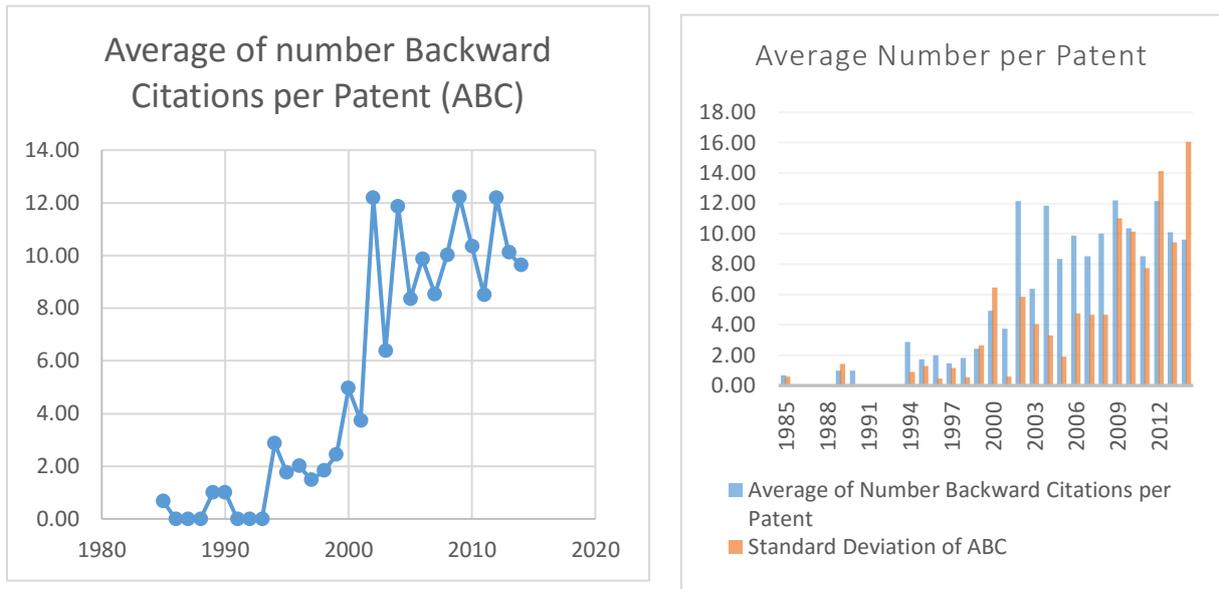

*Figure 17: Average number of backward citations and the Standard deviation*

## 5.2 Productivity improvement rate

The results on productivity improvement rate will be break down in three parts referring to the steps mentioned in the method section 4.1, i.e. trends obtained for 1996 to 2015 period using average values from state and patents data, trends from patents data using only the best yield reported in patents from 1985 up to 2010 and to 2015 and finally trends resulting from a difference-in-differences kind of analysis aiming at computing a more reliable and confident improvement rate by controlling for weather changes over time). The results of the performance trends are reported in Table 5 including the R² and the $P_{value}$.

First, we used annual time series data gathered from patents (70 hybrid corn not inbred) and state field-tests to fit the performance trends and get the improvement rate for hybrid corn. Figure 18 shows these trends for 20 years' period coverage (1996-2015) and the rate K is noted at the top left. Others detailed graphs obtain from the each of the 3 States data source are reported in appendix.

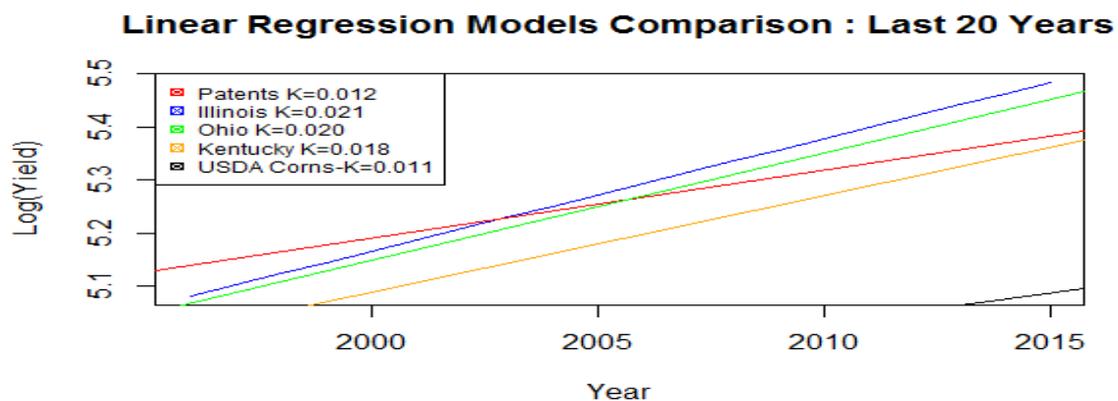

*Figure 18: Performance trends fitted from Patents and States data*

We have also calculated the improvement rate on corn productivity in the United States using USDA data from 1930 to 2015 and we found that, as shown on Figure 19, the linear regression fitted ($R^2 = 0.92, p\_value < 2.2 * 10^{-16}$) on the right plot gives roughly $0.023$ (which is 2.3% improvement per year and close to empirical rate on hybrid corn) as the rate of yield improvement in the corn domain in the US according to Moore's law. It should be emphasized that hybrid corn-similar to other technological domains- involves many technological changes over time.

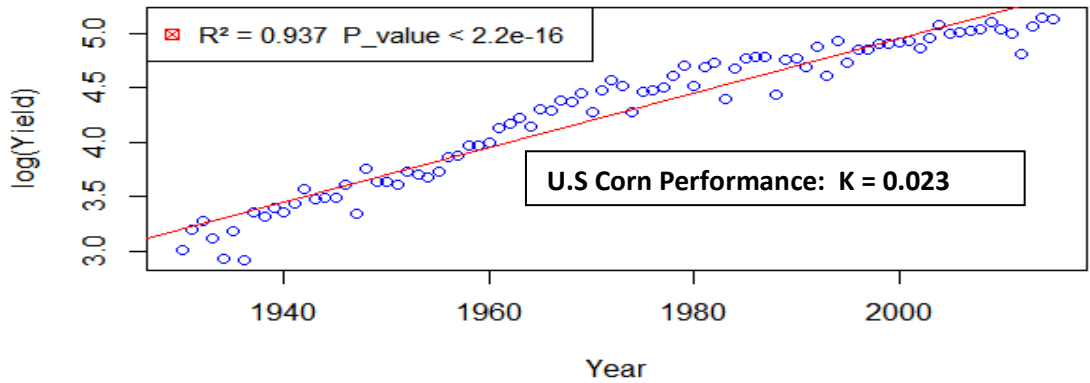

*Figure 19: United States corn productivity improvement from 1930 to 2015*

Secondly, the next set of rates comes from patents data only where we reported the best yield achieved for each year among our patent set and fit the improvement rate using Moore's law. Figure 19 below shows the trends obtained for hybrid corn patents filed from 1985 to 2010 and up to 2014 where the rates are respectively 1.5% and 1.22%. This shows that the drop in average yields observed for patented variety after 2010 does not strongly affect the estimation of the rate of improvement.

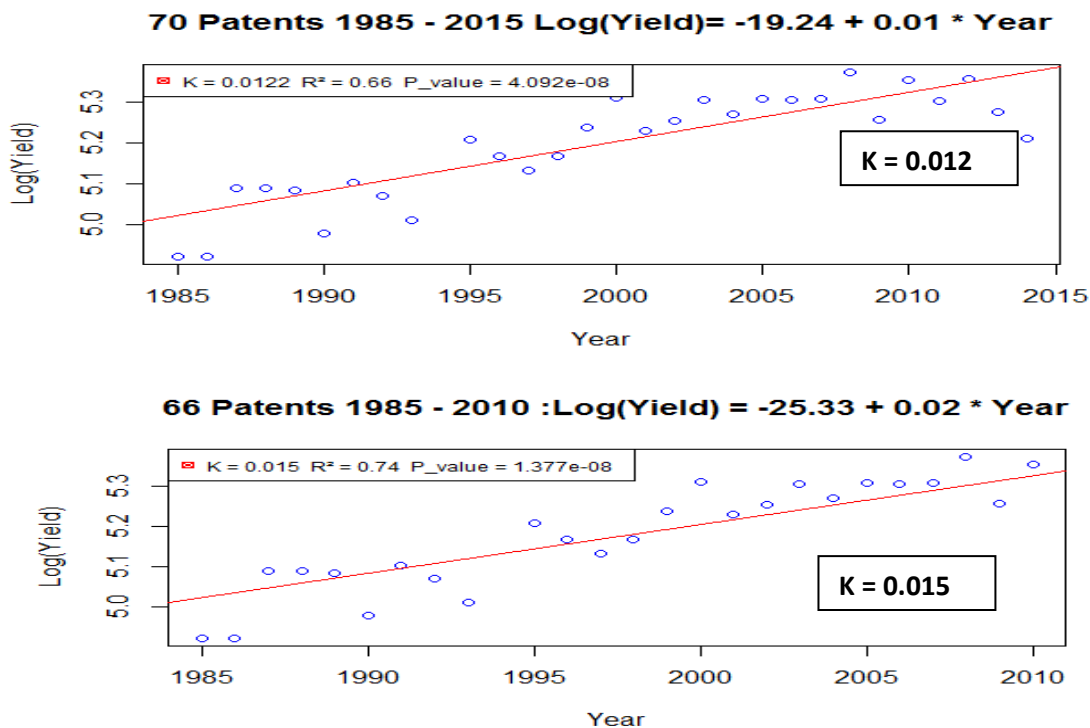

*Figure 20 : Hybrid corn performance trends fitted using patents*

Thirdly, as announced in the method section, we also investigated how the weather and soil variations affect the yield and the rate K calculate below using average yield across states regions. We implemented the method described in section 3.1 and found that in Illinois field-test, variety 'Whata 5656' and '4373' in North region and variety '33P67' in West Central region are tested across several years 7 or 8, in the same region. Then, by using these three varieties as controls, we fitted the trends of the best yield normalized by the yield of the control varieties. We found respectively 0.013, 0.02 and 0.029 for these three control varieties (improvement rate observed for 6 or 7 years as shown on Figure 20 and 21.) We noted that these improvement rates are between 1.3% and 2.9% per year, while we got 2.1% as average improvement rate in the whole state of Illinois. These findings can be regrouped in two points: Firstly, the average productivity values used to fit Illinois performance trend was a good estimation of the overall improvement rate observed in the State. Secondly, the improvement rate varies a lot from a region to another, 1.6 % difference observed between Illinois's North region and West central's with five years' overlap which is more than a factor of two. This relatively high difference between two regions of the same State tend to indicate that the Moore's law or the exponential models may not the best method to estimated improvement in hybrid corn domain or in food industry in general. Others models such as Wright's law, Goddard's law (drop in prices due to greater productivity), Nardhaus' Synthesis (combines elements of Moore's law and Wright's law) are interesting alternative models to try. Note that FAO (Food and Agricultural Organization of the United Nations) found that food production growth rates from 200 to 2013 vary from 1.03% in Europe to 3.49% in Africa [29]. These rates are relatively close to what we found as empirical improvement rate in hybrid corn domain which were between 1.2% and 2.4%. However, their result is based on Least-Squares method which might be better to estimated improvement rate in hybrid corn productivity than exponential fits.

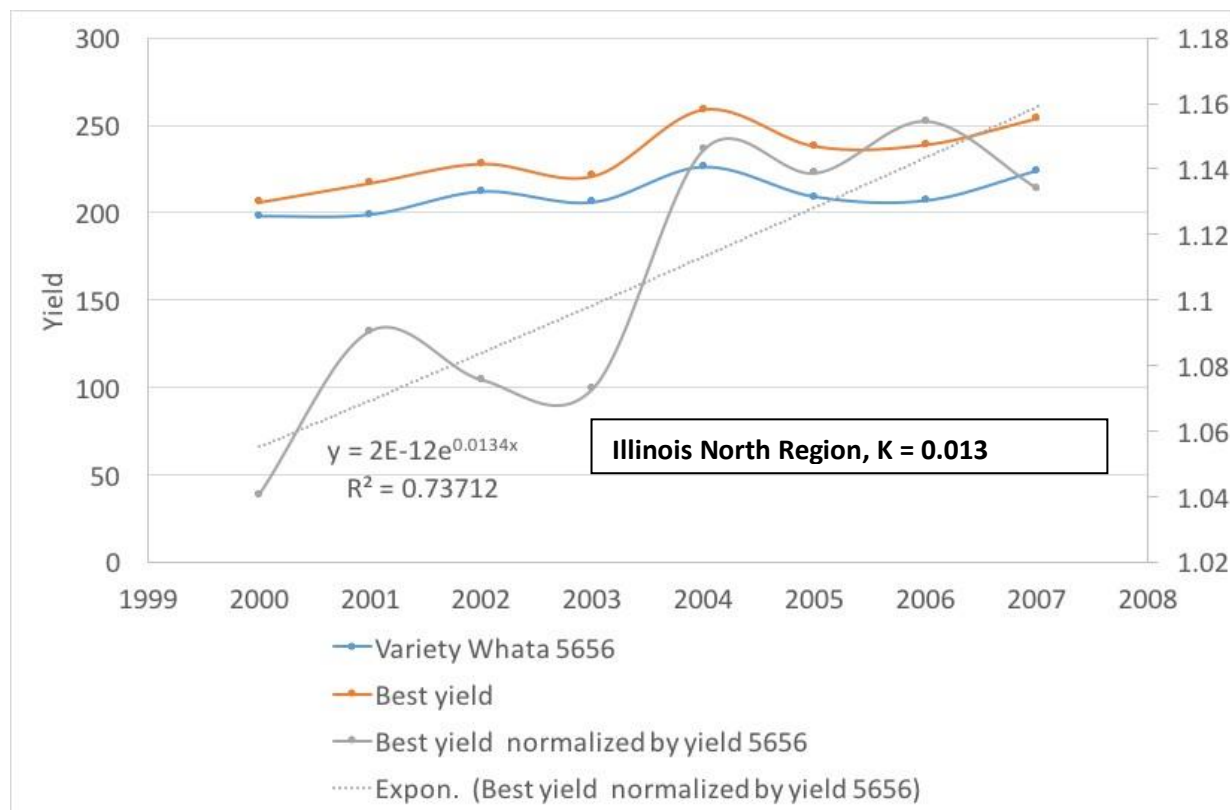

Figure 21: Illinois North Region Improvement rate corrected from weather and soil effects

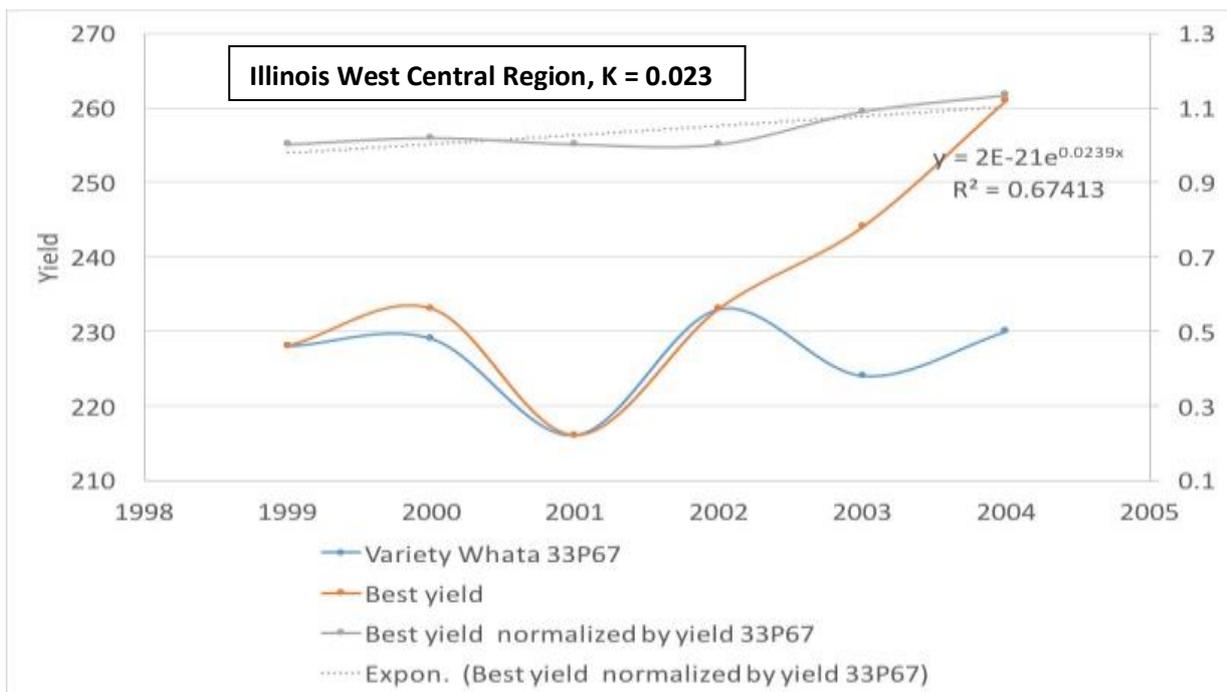

Figure 22: Illinois West central Region Improvement rate corrected from weather and soil effects

This empirical improvement rate in hybrid corn domain from Illinois regions cleaned from weather and soil effect is particularly relevant and represents a good estimation of improvement in corn domain in the United States since Illinois plays a major role in U.S. corn production. Indeed, Illinois is the second producer of corn production in the U.S. with more than 12 012 Million bushel produced in 2015 right behind Iowa[15] who produced around 12 505 Million bushels last year according to USDA (United States Department of Agriculture) [33]. These two majors corn producers in the US - the only ones that produced more than 200 Million these last years - are followed by Nebraska (1 692 Millions) and Minnesota (about 1 428 Million bushels) while the other states produced less than 1 000 Million bushel in 2015 [34].

We have sum up these empirical rates in Table 5 including the Pearson coefficient and the $P_{value}$. The overall results indicate that the models fitted using Moore's law is statistically consistent with high R² and low $P_{value}$. An important point to note is that all these empirical rates from different data source are close and between $0.012$ and $0.024,$ which makes the results more reliable. For example, for a sample of patents filed until 2010, we have observed a rate of 0.015 (1.5% of improvement per year) with $R^2 = 0.74$, $P_{value} = 1.37 * 10^{-8}$. Even if these rates are relatively low compared to other technological domains shown in Table 1 (for 28 domains), they are consistent with food productivity domain progress which is fairly slow compared to other domains.

---

[15] Iowa field-test data were provided from 2004 to 2015, a short time period compared to the 20 years data obtained on Illinois State filed-trials whose production is relatively close to Iowa's, thus, Illinois data was relevant data source

| DATA SOURCE | RATE K (%) | R² | P_VALUE |
|---|---|---|---|
| PATENTS_1996-2015 | **1.20** | 0.80 | $6.68 * 10^{-8}$ |
| OHIO_1996-2015 | **2** | 0.60 | $5.80 * 10^{-5}$ |
| KENTUCKY_1996-2015 | **1,8** | 0.39 | 0.003 |
| ILLINOIS_1996-2015 | **2.10** | 0.76 | $4.38 * 10^{-7}$ |
| PATENTS -1985-2010 | **1.50** | 0.74 | $4.092 * 10^{-8}$ |
| PATENTS – 1985-2014 | **1.22** | 0.66 | $1.337 * 10^{-8}$ |
| USDA_US CORN_1830-2016 | **2.48** | 0.93 | $2.2 * 10^{-16}$ |
| ILLINOIS_WEST CENTRAL REGION | **2.39** | 0.67 | |
| ILLINOIS_NORTH REGION | **1.34** | 0.73 | |

*Table 5: Empirical Improvement rates observed in Hybrid corn domain*

We can now compare these empirical improvement rates to the estimated rates calculated for several cases. Table 6 contains the improvement rates obtained using the two estimating models described above, for different dataset (i.e only hybrid or only inbred or both, filed up to 2005 and to 2013). We observed that only estimated rate using hybrid corn patents filed until 2005 is in accordance with observed values.

| Variety Type | HYBRID | | INBRED | | Hybrid + Inbred | |
|---|---|---|---|---|---|---|
| Filing Year up to | 2005 | 2013 | 2005 | 2013 | 2005 | 2013 |
| Benson & Magee | -0.0146 | 0.5116 | -0.0157 | 0.07803 | -0.0154 | 0.2486 |
| Triulzi & Magee | 0.015 | 28.4969 | 0.1658 | 0.26740 | 0.0428 | 0.1617 |

*Table 6: Estimated yearly improvement rate in Hybrid Corn domain*

Indeed, Table 6 shows an important difference in improvement rates for all these cases. For example, the predicted rate from Benson and Magee's model [17] are negative for all patent sets filed up to 2005 (around -0.015). The negative rates are due to a very low value of average year of patents publications in hybrid corn which impacted in the model equation. Although, these rates should be very surprising since the predicted rate for food engineering domain (in which hybrid corn is included) found by Benson and Magee (2016) was negative too : - *0.107*.

An interesting finding is that we got the much higher prediction for all the estimating models when patents beyond 2008 are included using both of the models. The timing indicates that the drastic in change in patenting practices shown in subsection 5.1.2 has led to these extremely high estimations compared to observed rates, except in one case. Indeed, the increase in backward citations started around 2008 will change forward citations for patents in the domain and thus Cite3 which a key variable in equation (4) and eventually the rate Z in equation (7) and the centrality C which are both linked to citations. Moreover, the other changes observed in the rise of number of filed patents will also impact on the average publication year (the second variable of equation (4) model) therefore it will affect the value of the estimated rates.

Note that, the higher and thus worst rate obtained using Triulzi and Magee's model indicates that the centrality and rate of impactful growth Z have considerably changed after patenting practices explosion occurs since the same model gives both the best and also the worst estimation of the existing improvement rate. Furthermore, it is important to point out that the predicting rate resulting from Triulzi and Magee 's model applied on hybrid varieties filed until 2005[16] (i.e which is cleaned from effects due to patenting practices changes - which is 0.015 - is in good agreement with the empirically observed rates (between 1.2% and 2.4%). This estimated rate is also very similar to the observed improvement rate obtained from patents data (which is 0.015 for hybrids patents filed until 2010) which means that the estimating model is valid when it is computed with correct variables.

## 5.3 Hybrid corn performance effect on patent citations

We fitted the models describe in Table 3 in section 4.3 and reported the results in Tables 7 (OLS Regression) and Table 8 (Poisson Regression) below. The significance code for the coefficients reported in these tables is: **\*\*\*p<0.001, \*\*_p<0.01_ \*p<0.05** and the standard error is note into brackets. As already mentioned above, the aim of this study is to investigate whether the best performing varieties receive more citations. We can note several points according to the statistical results. As expected, models with Poisson regression give a lower $P_{value}$ (i.e. are more significant) than the results obtained with OLS Regression since I is better to run Poisson regression for count data such as citations in patents. In the following comparison for the first 4 models, we will focus on Poisson regressions results since citations are count variables and may mention OLS regression result for additional comments.

| OLS MODEL | 1 | 2 | 3 | 4 |
|---|---|---|---|---|
| DEPENDENT VARIABLE | Forward Citations | Cite3 | Cite3_rank_percentile | Forward Citations |
| PERFORMANCE_RATIO | 30.553** (9.921) | -4.365 (3.908) | -0.029 (0.345) | 20.48 (11.98) |
| YEAR_FILED | 0.795*** (0.141) | 0.151** (0.054) | 0.196*** (0.004) | |
| INTERCEPT | -1614*** (289.9) | -296.382** (107.6) | -38.784*** (9.508) | -14.09 (12.46) |
| R² | 0.379 | 0.112 | 0.20 | 0.048 |

*Table 7: OLS Regression models results*

---

[16] No highly cited hybrid corn patent was found before 2005 therefore the parameter Z was equal to zero.

| POISSON MODEL | 1 | 2 | 3 | 4 |
|---|---|---|---|---|
| DEPENDENT VARIABLE | Forward Citations | Cite3 | Cite3_rank_percentile | Forward Citations |
| PERFORMANCE_RATIO | 3.540*** | -3.026 * | -0.137 | 2.0336*** |
|  | (0.488) | (1.204) | (1.770) | (0.474) |
| YEAR_FILED | 0.109*** | 0.817 *** | 0.039 |  |
|  | (0.009) | (0.0131) | (0..025) |  |
| INTERCEPT | -220*** | -159.70 | -79.623 | -0.154 |
|  | (19.64) | *** | (50.100) | (0.502) |
|  |  | (26.205) |  |  |
| AIC | 437 | 325.24 | INF | 562 |

*Table 8: Poisson Regression models results*

One of the implications of this statistical analysis is the apparent effect between performance achieved by a variety and the citations received by its patent. Indeed, the comparison of models 1, 2 and 3 indicates two points. First, the total number of received citations (until 2015) in patents is more correlated to the relative performance of the variety (lower $P_{value}$) than the two other citations variables, Cite3 (number of citations received within 3 years after patent publication) and cite3 rank percentile. The second point and also an answer to our second research question is that there is an apparent correlation between number of citations received by a patent and the performance of the patented variety according to both OLS and Poisson regression with Performance ratio and also Year as independent variables. Model 1 indicates that the correlation effect seems to be stronger in the Poisson regression (which make sense due to the numeric character of the citations data) with a $P_{value}$ lower than $4.33 * 10^{-13}$ - for all the independent variables coefficients, including the intercept - than in the OLS regression where found $R^2 = 0.379$ and overall value of the test of $1.217 * 10^{-06}$ with a value of 0.00319 for the performance ratio coefficient.

In particular, the regressions run tend to confirm what Moser et al (2015) [22] found in their research where they found a correlation between improvement in yield and citations received (model 1). Indeed, they got for improvement in yield[17] variable (we computed that variable as the ratio of patented variety yield over comparison variety yield) a $P_{value}$ < 0.05 and R² = 0.554 for OLS Regression. Our research gives a lower $P_{value}$ of 0.00319 and R² of 0.0379 for the OLS regression. And for the Poisson regression they found a $P_{value}$< 0.01 for improvement in yield variable while we got a $P_{value}$ of $4.33 * 10^{-13}$ for the performance ratio variable, assimilated as the relative performance of the variety and similar to the improvement in yield of the patented variety. To sum up, the different models fitted, on a basis of 70 patents dataset indicate that there is correlation between the performance ratio of the variety and the number of total forward citations received by the patent, even if the correlation is not linear as we can see it in Figure 23 the citations against performance ratio.

---

[17] The improvement in yield, in Moser et al paper, was calculated as the difference between the yield of the patented hybrid and the yield of the highest-yielding existing hybrid in the same trial, divided by the yield of the highest-yielding existing varie12ty. Their data include 269 individual patents (315 patent-hybrid pairs). See [22], pages 3-4.

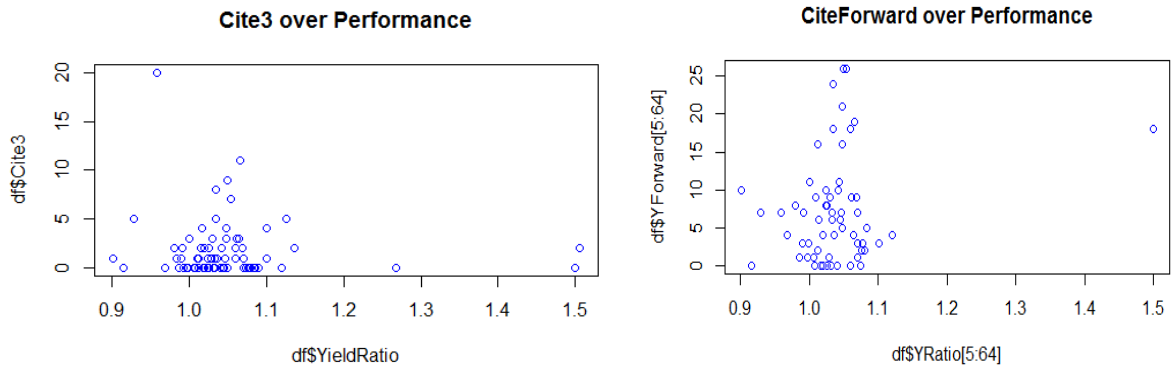

*Figure 23 : Forward Citations over Performance ratio*

We have run a negative binomial regression on these four models and the results confirm an effect between these variables. These results are reported in the appendix.

The model 4, which analyzes the effect of performance ratio on patents total received citations, even without the patent filing year as control parameter indicates a correlation between these two variables as well using Poisson regression ($p_{value} = 1.85 * 10^{-5}$). Although, that effect is not linear at all with a very poor R² = 0.048 in contrast to 0.37 for linear model with the year fixed effect taken into account in model 1. This poor linear effect, as shown on figure 23 makes sense since the forward citations are a type of count data and year effect has not been taken into account.

## 5.4 Limitations

Our research provides interesting results on hybrid corn domain performance trends in a macro analysis and further statistical analysis at the individual seed level. However, it is important to point out the limitations of the research. First, we have noticed a relatively high noise-signal ratio in the average yield values used to calculate improvement rate in states. Note that, empirical rate from patent data are not subject to the noise observed in states average values because instead of average yield value, we used the best yield achieved found in patents, for each year. Moreover, the Moore's law or exponential fit might not be the best way to estimate improvement rate in hybrid corn, since food industry domain is known for his relatively slow progress and innovation compared to others technological domains. It would be interesting to compute the empirical data using others forecasting models such as Wright's law or others linear models with additional parameters including the conditions, quality of soil to compare them and deduce the best estimating one.

Secondly, an important point on the limitations of this research in general is the reliability of citations number received by a patent in hybrid corn domain which impacted on the performance of the two estimating models. Indeed, the citations has been used as dependent variable to investigate on the second research question but have also represented a key metric in both of the performance predicting models (cite3 and centrality depends somehow on citations). Due to the few number of patenting firms, hybrid corns patent is more likely to be cited by patent controlled by the same firm that holds the patent in opposite in others domains. For example, in Kentucky field-test patented varieties from 1998 to 2005, 84% of received citations by Pioneer patents were actually made by other Pioneer patents which indicates that *self-citation* is current in hybrid corn or seed industry. Since the estimating models have shown to be controversial in hybrid corn domain (but relevant for 28 technological domains tested), the models could be improved with additional parameters which will allow to take into account change in patenting practices and provide more accurate results on the rate of improvement in that domain, cleaned from any notable effect.

Furthermore, our dataset of 70 patents is not be very sufficient to indicate overall results on hybrid corn domain, especially with existing genetic stack[18] effects. However, the findings on the correlation between citations and performance ratio at the micro level, tested over different regressions are interesting and tend to confirm the results presented in *Moser et al (2015)*. Note that the empirical rates of improvement in the corn productivity found has shown statistical robustness from several data sources which indicates reliable results on this macro-research question

# 6. Conclusion

This paper presents a significant result on productivity improvement rate from hybrid corns, an important sub-domain of food production. It is important to note that, by several different measures – from patents, field-test data and a forecasting model – we found improvement rates reasonably close and between 1.2% and 2.4% per year. Moreover, the regressions fitted have shown a statistical robustness where the strongest forecasting model is obtained from a sample of patents granted between 1996 and 2015 ($R^2 = 0.80, P_{value} = 6.68 * 10^{-8}$). We also noticed that the growth rate in the hybrid corn domain – like in food engineering in general - is much lower than other technological domains, despite this low rate, it seems to be sufficient to support a high level food supply to overtake population growth so far. However, FAO keeps sounding the alarm about food insecurity and the substantial insufficiency in food supply in some developing regions. Furthermore, the performance trends analysis does not support a hypothesis that innovation and improvement in biotechnology is already leading to a higher improvement rate despite hybrid crops wide adoption in agricultural domain.

One main point of the findings is the change in patenting practices, in particular the important increase in number of patents applications and drastic rise in average number of backward citations per patent which occurred after 2000 in hybrid corn domain. It seems that these changes have considerably impacted the parameters of the two forecasting models tested, which led to a breakdown in the results. It can be interesting to investigate on how to correct these variables from changes effect to better adapt them to hybrid corn domain, since the two models have shown strong and confident results when applied to other domains.

Finally, focusing at the seed level through the individuals 70 hybrid corn patents (granted between 1986 and 2015), regressions models results tend to confirm a correlation between the seed performance and the total citations received by the corresponding patent, as first shown in a NBER (National Bureau of Economic Research) working paper by *Moser et al (2015)* in which they work with 269 hybrid corn patents (granted between 1986 and 2005). Our findings can be more accurate if the study is extended to a bigger patent dataset.

---

[18] Gene stacking – also known as gene pyramiding – is the process of including more than one transgenic event in one plant to produce stacked traits, stacked transformation events, or a stacked genetically modified organisms (GMO) [37]
According to a Mother Jones article [38] "Stacked traits" tend to achieve lower yield than other Genetically Modified varieties.

A further study that could follow up on this work, could be to investigate on other performance predicting models or develop additional parameters on existing models to better understand performance improvement rate on hybrid corn domain and food industry in general.

# References

1. Malthus, RM. 1798. An Essay on the Principle of Population As It Affects the Future Improvement of Society, with Remarks on the Speculations of Mr. Godwin, M. Condorcet, and Other Writers. London, J. Johnson, in St. Paul's Church-yard

2. Dyson, T. 1996. Population and Food, Global trends and future prospects. Routledge Edition

3. Simon, JL. 1981. "The ultimate Resource " book (p. 348) Princeton University Press.

4. United Nations Environment Program. October 2012. Climate change, Resource efficiency, Ecosystem management. Growing Greenhouse Gas Emissions Due to Meat Production. Website consulted on August 12, 2016.

5. Brown, LF. 1991. 'What Food Indicators Say,' State of the World 1991, World Watch Institute, pages. 11-15

6. Evenson, R.E., D. Gollin. 2003. Assessing the impact of the green revolution, 1960 to 2000. Science 300: 758–762.

7. Kristen Michaelis on Food Renegate. Hybrid seeds vs Genetically Modified Organisms. Website consulted in August 14, 2016.

8. Steve Savage. June 16, 2016. Biotech opponents claim Monsanto 'controls' the world food supply? What are the facts? Website visited in August 12, 2016.

9. National Corn Growers Association. Data source: USDA, FAS Grain : World Markets and Trade, January 12, 2016. http://www.worldofcorn.com/#world-corn-consumption . Website visited on August 12, 2016.

10. National Corn Growers Association. Data source: USDA, FAS Grain : World Markets and Trade, January 12, 2016. http://www.worldofcorn.com/#world-corn-production . Website visited on August 12, 2016.

11. United States Department of Agriculture. Economic Research Service. Data on Feed Grains: Yearbook Tables. Last update on August 16, 2016. Data downloaded on June 17, 2016.

12. Moore GE. 1965. Cramming more components onto integrated circuits. Electronics Magazine.

13. Nagy B, Farmer JD, Bui QM, Trancik JE. 2013. Statistical Basis for Predicting Technological Progress. PLoS ONE. 8(2), 1–7. doi: 10.1371/journal.pone.0052669

14. Martino J. 1976. Examples of Technological Trend Forecasting for Research and Development Planning. Technological Forecasting and Social Change. 2 (3/4). 247–260

15. Koh H, Magee CL. 2006. A functional approach for studying technological progress: Application to information technology. Technological Forecasting and Social Change. 73(9), 1061–1083. doi: 10.1016/j.techfore.2006.06.001

16. Fleming L. 2001. Recombinant Uncertainty in Technological Search. Management Science. 47(1), 117–132. doi: 10.1287/mnsc.47.1.117.10671

17. Benson CL, Magee CL. 2015. Quantitative Determination of Technological Improvement from Patent Data. PlosS ONE 10(4) : e0121635 doi : 10.1371/journal.pone.0121635

18. Magee et al. 2016. Quantitative empirical trends in technical performance. Technological Forecasting and Social Change. http://dx.doi.org/10.1016/j.techfore.2015.12.011

19. Benson CL, Magee CL. 2014. Using Enhanced Patent Data for Future-Oriented Technology Analysis, 5th International Conference on Future-Oriented Technology analysis (FTA) – Brussels, 27-28 November 2014

20. Benson, CL, Magee, CL. (2013). A hybrid keyword and patent class methodology for selecting relevant sets of patents for a technological field. Scientometrics, 96(1), 69–82. doi:10.1007/s11192-012-0930-3

21. Triulzi G, Magee CL. 2016. Predicting Technology Performance by Mining Patent Data. Presentation at ISS Conference in Montreal on July 8, 2016.

22. Moser et al. 2015. Patent Citations and the Size of the Inventive Step - Evidence from Hybrid Corn. National Bureau of Economic Research.

23. Cargill. The importance of innovation to food security. Data source: USDA, Cargill analysis. Website consulted on August 14, 2016.

24. Patentswiew queries guide on http://www.patentsview.org/api/uspc.html. Data on patents having Hybrid Corn or Inbred Corn or Hybrid Maize or Inbred Maize in there title or abstract, filed from 1985 : Query

25. Ohio Crop Performance Trials Archive. Data on http://hostedweb.cfaes.ohio-state.edu/perf/archive.htm downloaded on July 25, 2016.

26. Kentucky Crop Performance Trials Archive. Data on http://www2.ca.uky.edu/cornvarietytest/Archives.html downloaded on June 3rd 2016.

27. Illinois Crop Performance Trials Archive. Data on http://vt.cropsci.illinois.edu/corn.html downloaded on July 9, 2016

28. United States Department of Agriculture. Economic Research Service. Data on Yield average annual time series data. Feed grain database. Visited on July 18, 2016.

29. Food and Agricultural Organization of the United Nations. Statistics Division. Production growth rate by region from 2000 to 2013.

30. GMO Genetic Literacy Project. Do Monsanto and Big Ag control crop research and world food supply? Article download on August 14, 2016.

31. Agweb. 2015 Seed family tree. Farm Journal – 57. Summer 2015. Website visited on August 14, 2016.

32. Tom Philpott. December 14th 2015. Move Over, Monsanto: The Pesticide and GMO Seed Industry Just Spawned a New Behemoth. Mother Jones article. Visited on August 10, 2016.

33. World of Corn. U.S Production per state from USDA data. http://www.worldofcorn.com/#us-corn-production-by-state. Visited on August 15th 2016.

34. Rob Cook. 2015. States that produce the most corn on beef2live. Visited on August 15, 2016.

35. Wikipedia. Land grant university. Last update on July 24, 2016. Visited on August 7, 2016.

36. Wikipedia. DuPont. Last update on August 18, 2016. Visited on August 15, 2016.

37. Biosencience for farming in Africa. What is Genetic stackes ? http://b4fa.org/bioscience-in-brief/new-plant-technologies/what-is-gene-stacking/. Visited on 10 August, 2016.

38. Tom Philpott. February 13th 2013. Do GMOs really have Higher Yield ? Mother Jones article. Website visited on August 10, 2016.

39. Luther Burbank. 1926. Current Challenges and Past Progress for Food Supply and Quality. Website visited on August 7, 2016.

40. Cornell University. Northeast Region Certified Crop Adviser (NRCCA) Study Resources. What is a Hybrid ? Website visited on August 7, 2016.

# Appendix

*Figure 24: Empirical Performance trends for Illinois, Kentucky and Ohio*

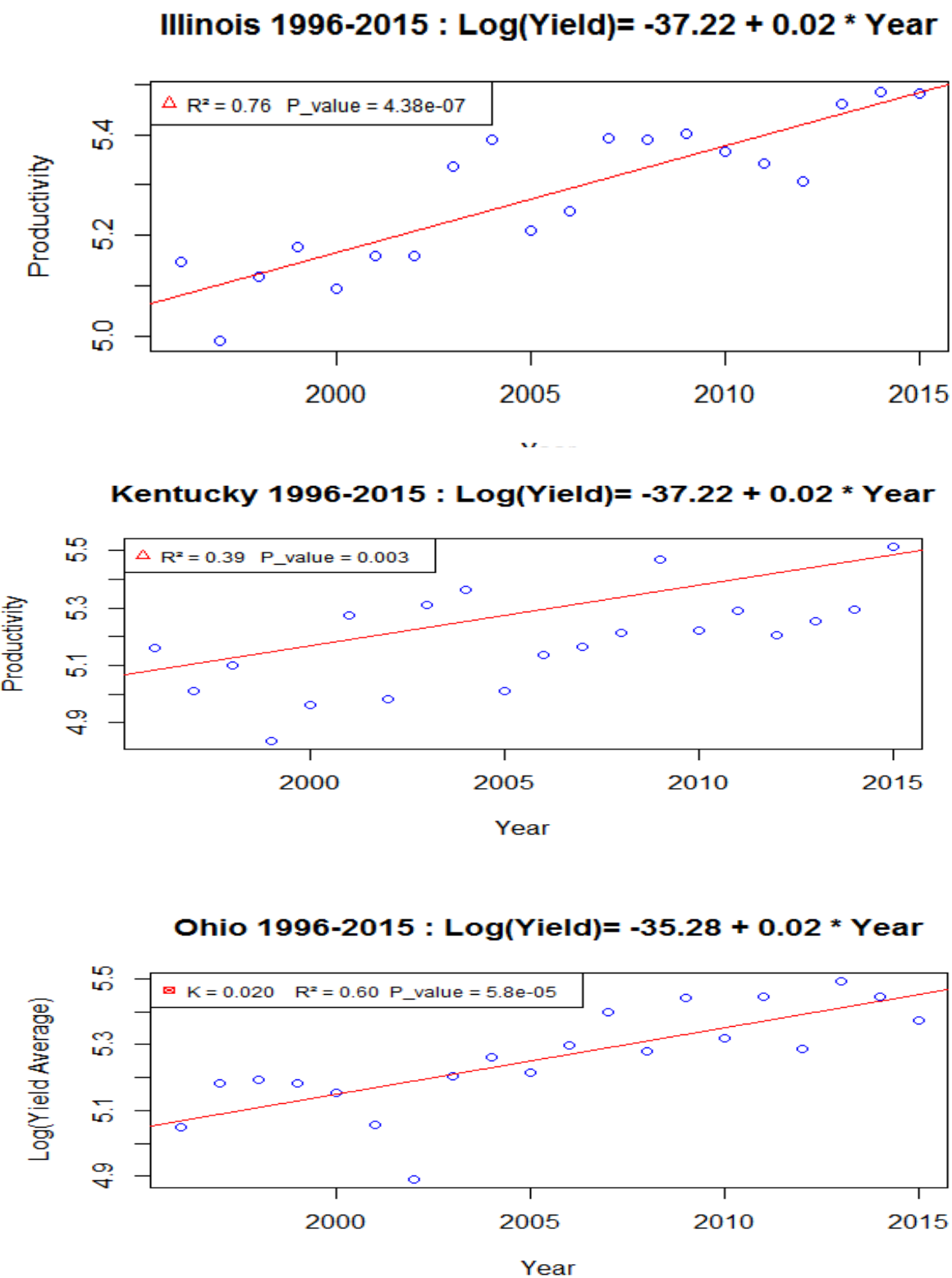

*Figure 25: Negative Binomial Regression Results*

The Model 1 to 4 refers to what is described in the Method section.

| IV\MODEL | 1_BC | 2_ | 3_CRP | 4_BC |
|---|---|---|---|---|
| Performance_Ratio | 3.231** | 1.185 | 0.198 | 2.0489 |
| Year | 0.097*** | 0.164*** | 0.094** | |
| Yield | | | | |
| Intercept | -195.44*** | -330.45*** | -189.70** | -0.2195 |
| Std. Error | 0.432 | 0.934 | 1126086 | 0.25 |
| AIC | 339.03 | 182.85 | 86.40 | 351.62 |
| 2xLog(likelihood) | -331.03 | -173.845 | -78.40 | -345.62 |

*Figure 26: Code used to clean patent data and Extract Variety Name from patent title*

```python
#!/usr/bin/env python
# -*- coding: utf-8 -*-
import csv

## Code to clean patentsview.org data and extract the variety

# Lists that will contain the values
title = []
year = []
patent = []
claim = []
citation = []
inventor = []
brand = []

## Reading the data from the csv retuned using pandas
with open('patent_Hybrid_Inbred_Corn_Maize_DATA.csv') as csvfile:
    reader = csv.reader(csvfile, delimiter = ';')
    for row in reader:
        patent.append(row[1])
        year.append(row[2])
        title.append(row[3])
        brand.append(row[4])
        citation.append(row[5])
        claim.append(row[6])
        inventor.append(row[7])

## To extract only Brand name and delete the other caractèrs around
brands = []
brand_types = [ '[{\'assignee_organization\': \'' ,
'\'}', ' {\'assignee_organization\': \'' , '\'}]',
'\"}]', '[{\'assignee_organization\': \"']
for b in brand:
    v = b
    for var in brand_types:
        v = v.replace(var, '')
    brands.append(v)

## Recuperation of on the year in the patent date
years = []
years.append(year[0])
for y in year[1:]:
    years.append(y[6:10])
```

```python
## Titles process to extract the hybr:
# Various possible begining of titles
a = 'Inbred corn line '
b = 'Hybrid corn plant and seed '
c = 'Methods and compositions of a hy
d = 'Hybrid genetic complement and co
e = 'F, hybrids and method for produc
f = 'Inbred maize line '
g = 'Hybrid maize plant and seed '
h = 'Hybrid maize plant & seed '
i = 'Synthetic corn hybrid '
j = 'Plants and seeds of hybrid corn
k = 'Maize hybrid variety '
l = 'Corn hybrid variety '
m = 'Hybrid maize variety '
n = 'Hybrid corn variety '
o = 'Maize variety '
p = 'Maize variety inbred '
q = 'Corn variety '
r = 'Corn variety inbred '
s = 'Inbred maize variety '
t = 'Inbred corn variety '
u = 'Maize variety hybrid '
v = 'Corn variety hybrid '
w = 'Maize hybrid '
x = 'Maize inbred '
y = 'Corn inbred '
z = 'inbred '
zz = 'hybrid '
aa = 'Inbred sweet corn line '
bb = 'Inbred maize seed and plant '
cc = 'Hybrid maize seed and plant '
ww = 'Hybrid corn line '
xx = 'Hybrid maize '
yy = 'Corn '
zzz = 'synthetic population '
aaa = 'Hybrid maize plant & seed'
bbb = 'Hybrid maize plant and Seed '
ccc = 'Inbred maize lind '
ddd = 'Hybrid maize plant and seeds '
eee = 'Inbred maize lines '
fff = '& seed '
ggg = 'Imbred corn line '
hhh = 'plant and seeds '
iii = 'Inbred Variety '
jjj = 'Inbred com line '
kkk = 'Variety corn line '
lll = 'aize variety '
mmm = 'variety '
nnn = 'Inbred line '
ooo = ' and seeds thereof'
ppp = 'Inbred corn plants '
qqq = 'designated '

variety = []
types = [a,b,c,d,e,f,g,h,i,j,k,l,m,n
cc,dd,ee,ff,gg,hh,ii,jj,kk,ll,mm,nn,
aaa,bbb,ccc,ddd,eee,ddd,fff,ggg, hhh

for t in title:
    v = t
    for var in types:
        v = v.replace(var, '') # We
    variety.append(v)

ggg = 'Imbred corn line '
```